\documentclass[traditabstract]{aa}
\bibpunct{(}{)}{;}{a}{}{,} 
\usepackage{graphicx}
\usepackage{mathrsfs}
\usepackage[varg]{txfonts}
\usepackage[T1]{fontenc}
\usepackage{color}
\usepackage{stfloats}

\begin{document}

\title{Size and kinematics of the low-ionization broad emission line region from microlensing-induced line profile distortions in gravitationally lensed quasars}
\author{Damien Hutsemékers\inst{1,}\thanks{Research Director F.R.S.-FNRS} 
   \and Dominique Sluse\inst{1}
   \and Đorđe Savić\inst{1}
       }
\institute{
    Institut d'Astrophysique et de G\'eophysique,
    Universit\'e de Li\`ege, All\'ee du 6 Ao\^ut 19c, B5c,
    4000 Li\`ege, Belgium
    }
%
%
\titlerunning{Size and kinematics of the quasar low-ionization broad emission line region from microlensing} 
\authorrunning{D. Hutsem\'ekers et al.}
\abstract{
Microlensing-induced distortions of broad emission line profiles observed in the spectra of gravitationally lensed quasars can be used to probe the size, geometry, and kinematics of the broad-line region (BLR). To this end, single-epoch \ion{Mg}{ii} or H$\alpha$ line profile distortions observed in five gravitationally lensed quasars, J1131-1231, J1226-0006, J1355-2257, J1339+1310, and HE0435-1223, have been compared with simulated ones. The simulations are based on three BLR models, a Keplerian disk (KD), an equatorial wind (EW), and a polar wind (PW), with different sizes, inclinations, and emissivities. The models that best reproduce the observed line profile distortions were identified using a Bayesian probabilistic approach. We find that the wide variety of observed line profile distortions can be reproduced with microlensing-induced distortions of line profiles generated by our BLR models. For J1131, J1226, and HE0435, the most likely model for the \ion{Mg}{ii} and H$\alpha$ BLRs is either KD or EW, depending on the orientation of the magnification map with respect to the BLR axis. This shows that the line profile distortions depend on the position and orientation of the isovelocity parts of the BLR with respect to the caustic network, and not only on their different effective sizes. For the \ion{Mg}{ii} BLRs in J1355 and J1339, the EW model is preferred. For all objects, the PW model has a lower probability.  As for the high-ionization \ion{C}{iv} BLR, we conclude that disk geometries with kinematics dominated by either Keplerian rotation or equatorial outflow best reproduce the microlensing effects on the low-ionization \ion{Mg}{ii} and H$\alpha$ emission line profiles. The half-light radii of the \ion{Mg}{ii} and H$\alpha$ BLRs are measured in the range of 3 to 25 light-days. We also confirm that the size of the region emitting the low-ionization lines is larger than the region emitting the high-ionization lines, with a factor of four measured between the sizes of the \ion{Mg}{ii} and \ion{C}{iv} emitting regions in J1339. Unexpectedly, the microlensing BLR radii of the \ion{Mg}{ii} and H$\alpha$ BLRs are found to be systematically below the radius-luminosity ($R -L$) relations derived from reverberation mapping, confirming that the intrinsic dispersion of the BLR radii with respect to the $R-L$ relations is large, but also revealing a  selection bias that affects microlensing-based BLR size measurements. This bias arises from the fact that, if microlensing-induced line profile distortions are observed in a lensed quasar, the BLR radius should be comparable to the microlensing Einstein radius, which varies only weakly with typical lens and source redshifts.
}
\maketitle
%
%
%

\section{Introduction}
\label{sec:intro} 

Microlensing-induced distortions of broad emission line (BEL) profiles are commonly observed in the spectra of some gravitationally lensed quasar images, most often as red--blue or wings--core distortions \citep{2004Richards, 2004Metcalf, 2006Keeton, 2007Sluse, 2011ODowd, 2011Sluse, 2012Sluse, 2013Guerras, 2014Braibant, 2016Braibant, 2016Goicoechea, 2017Motta, 2018Fiana, 2020Popovic, 2021Fian}.   They can be interpreted in terms of the differential magnification of spatially and kinematically separated subregions of the broad emission line region (BLR). Various effects have been predicted depending on the BLR models \citep{1988Nemiroff, 1990Schneider, 2001Popovic, 2002Abajas, 2007Abajas, 2004Lewis, 2011ODowd, 2011Garsden, 2014Simic, 2017Braibant}, suggesting that microlensing could provide constraints on the size and kinematics of the BLR, as was first shown by \citet{2005Wayth}, \citet{2011Sluse}, \citet{2011ODowd}, and \citet{2013Guerras} (for a more comprehensive overview, see \citealt{2024Hutsemekers}).  

By comparing simulated line profile distortions to observed ones following the method developed in \citet{2017Braibant} and \citet{2019Hutsemekers}, we found that the \ion{C}{iv} line profile distortions observed in four lensed quasars can be reproduced with typical BLR models. Flattened, disk-like, geometries were found to best represent the BLR \citep{2021Hutsemekers, 2023Hutsemekers, 2024Hutsemekers, 2024Savic}. The size of the \ion{C}{iv} BLR was measured and found to follow the  radius-luminosity ($R-L$) relation from reverberation mapping, with possible evidence for the microlensing sizes lying, on average, below the $R-L$ relation \citep{2024Hutsemekers}. 

In this paper, we extend our analysis to the low-ionization BLR by investigating microlensing-induced line profile distortions observed in \ion{Mg}{ii} and H$\alpha$, using single-epoch spectroscopic data of five gravitationally lensed quasars. In Sect.~\ref{sec:data}, we describe the targets and the spectroscopic datasets we used. In Sect.~\ref{sec:bels}, we provide a detailed account of the microlensing distortions observed in BELs, with quantitative measurements. Section~\ref{sec:models} summarizes the method used to simulate microlensing line distortions based on representative BLR models. The results, including the determination of the most probable BLR models, the estimation of the BLR size, and the comparison with $R-L$ relations from reverberation mapping, are discussed in Sect.~\ref{sec:results}. Conclusions form the last section.

\section{Targets and data}
\label{sec:data}

We considered lensed quasars from \citet{2012Sluse} that exhibit clear line profile distortions due to microlensing in the \ion{Mg}{ii}~$\lambda$2800 or H$\alpha$ emission lines, and that were observed with a good-enough signal-to-noise ratio (S/N) to carry out our analysis (typically S/N $\gtrsim$ 30). These quasars are 1RXS~J113155.4$-$123155 (hereafter J1131), SDSS~J122608.02$-$000602.2 (J1226), and CTQ0327 aka Q1355$-$2257 (J1355).  We also considered HE~0435$-$1223 (HE0435), updating the analysis of \citet{2019Hutsemekers}, and SDSS~J133907.13$+$131039.6 (J1339), in which only the \ion{C}{iv} BLR microlensing was investigated by \citet{2024Hutsemekers}.

J1131 is a quadruply imaged quasar with an Einstein ring. The source is at redshift $z_s$ = 0.654 and the lens at redshift $z_l$ = 0.295 \citep{2003Sluse}. Long-slit spectra of images A, B, and C were obtained in the visible with the Very Large Telescope (VLT) equipped with the FOcal Reducer and low dispersion Spectrograph (FORS)2 on April 26, 2003, and in the near-infrared with the Infrared Spectrometer And Array Camera (ISAAC) on April 13, 2003. At each epoch, the spectra of A, B, and C were recorded simultaneously. The full spectral range includes the \ion{Mg}{ii} and H$\alpha$ lines. H$\beta$ is also present in the spectral range, but it was not used due to possible artifacts in the H$\beta$$+$[\ion{O}{iii}] region that makes the \ion{Fe}{ii} subtraction unreliable. Details of the observations and data reduction can be found in \citet{2007Sluse}.
 
J1226 and J1355 are doubly imaged quasars. For J1226, $z_s$ = 1.123 and $z_l$ = 0.517 \citep{2008Inada}, while for J1355, $z_s$ = 1.370 and $z_l$ = 0.701 \citep{2003Morgan,2006Eigenbrod}. Spectra of images A and B containing the \ion{Mg}{ii} line were obtained simultaneously with the VLT equipped with FORS1 on May 16, 2005, for J1226, and March 5 and 20, 2005, for J1355 \citep[see][for details]{2012Sluse}.

HE0435 is a quadruply lensed quasar with $z_s$ = 1.693 and $z_l$ = 0.454 \citep{2002Wisotzki,2012Sluse}. Since the \ion{Mg}{ii} line profile is contaminated by atmospheric absorption, we analyzed only the H$\alpha$ line, which was observed in the near-infrared.  Spectra of the four images were secured between October 19 and December 15, 2009,  with the VLT equipped with the Spectrograph for INtegral Field Observations in the Near Infrared (SINFONI) \citep[see][for details]{2014Braibant}.

Finally, J1339 is a doubly imaged quasar with $z_s$ = 2.231 and $z_l$ = 0.609 \citep{2014Shalyapin}. Spectra of images A and B covering the \ion{Mg}{ii} line were obtained on March 27, 2014, with the Gran Telescopio Canarias (GTC) equipped with the Optical System for Imaging and low-Intermediate-Resolution Integrated Spectroscopy (OSIRIS), and on April 6, 2017, with the VLT equipped with Xshooter \citep{2021Shalyapin}. They are publicly available from the GLENDAMA archive\footnote{\tt https://grupos.unican.es/glendama/database/} \citep{2018GilMerino}. Since the red wing of the line is cut off at the end of the 2014 spectrum, we considered only the \ion{Mg}{ii} line observed in 2017. H$\beta$ is also present in the 2017 Xshooter dataset, but the quality of the image B spectrum was not sufficient for our analysis.

\section{Broad emission line microlensing}
\label{sec:bels}

\begin{table*}[b]
\caption{Measured magnification and distortion indices.}
\label{tab:indices}
\renewcommand{\arraystretch}{1.2}
\centering
\begin{tabular}{lccccccc}
\hline\hline
 Object & I$_{\text{1}}$/I$_{\text{2}}$ & Line &  $M$ & $\mu^{cont}$ & $\mu^{BLR}$ & WCI & RBI \\
\hline
J1131  & A/B & \ion{Mg}{ii} & 1.80$\pm$0.17  &  0.32$\pm$0.03  &  0.87$\pm$0.08 &  0.74$\pm$0.02  & $+$0.066$\pm$0.012 \\
J1131  & A/B & H$\alpha$    & 1.80$\pm$0.17  &  0.37$\pm$0.05  &  0.84$\pm$0.08 &  0.83$\pm$0.02  & $+$0.026$\pm$0.012 \\
J1226  & A/B & \ion{Mg}{ii} & 1.14$\pm$0.10  &  2.09$\pm$0.19  &  1.47$\pm$0.13 &  1.41$\pm$0.04  & $-$0.108$\pm$0.012 \\
J1355  & A/B & \ion{Mg}{ii} & 2.95$\pm$0.25  &  1.64$\pm$0.14  &  1.14$\pm$0.10 &  1.12$\pm$0.03  & $+$0.120$\pm$0.011 \\
J1339  & B/A & \ion{Mg}{ii} & 0.26$\pm$0.04  &  4.74$\pm$0.74  &  1.15$\pm$0.18 &  1.03$\pm$0.11  & $+$0.146$\pm$0.040 \\
HE0435 & D/B & H$\alpha$    & 0.67$\pm$0.05  &  1.07$\pm$0.15  &  1.04$\pm$0.08 &  1.01$\pm$0.04  & $+$0.122$\pm$0.012 \\
\hline
\end{tabular}
\end{table*}

To characterize the line profile distortions induced by microlensing, we used the magnification profile $\mu(v)$,
\begin{equation}
\mu \, (v) =  \frac{1}{M}  \frac{F^l_{\text{1}} \, (v) }{F^l_{\text{2}} \, (v)} \; ,
\label{eq:muv}
\end{equation}
where $F^l_{\text{1}}$ and $F^l_{\text{2}}$ are the continuum-subtracted emission line flux densities simultaneously measured in images 1 (I$_{\text{1}}$, microlensed) and 2 (I$_{\text{2}}$, not microlensed), $M = M_{\text{1}} / M_{\text{2}}$ is the macro-magnification ratio of images 1~and~2, and $v$ the velocity computed from the line center at the source redshift. We also considered three indices integrated over the line or the magnification profiles (see \citealt{2017Braibant} or \citealt{2021Hutsemekers} for exact definitions): (1)~$\mu^{BLR}$, the total magnification of the line, (2) the wings--core index (WCI), which indicates whether the whole emission line is, on average, more or less magnified than its center, and  (3), the red--blue index (RBI), which measures the asymmetry of the magnification profile. A fourth index, $\mu^{cont}$, gives the magnification of the continuum underlying the emission line. $\mu(v)$ and the indices are independent of quasar intrinsic variations that occur on timescales longer than the time delay between the two images.  If the time delay is shorter than 40-50 days, \citet{2012Sluse} showed that the line profile distortions can be safely attributed to microlensing rather than to intrinsic variations seen with a delay between the images. 

An accurate measurement of the macro-magnification ratio, $M$, is needed to correctly estimate $\mu(v)$, $\mu^{cont}$, and $\mu^{BLR}$ (RBI and WCI are independent of $M$). Moreover, in Eq.~\ref{eq:muv}, $F^l_{\text{1}}$ and $F^l_{\text{2}}$ should be corrected for the differential extinction between images 1 and 2 that may arise from the different light paths through the lens galaxy. Since extinction equally affects the continuum and the lines, as opposed to microlensing, it can be incorporated into the factor, $M$, instead of correcting $F^l_{\text{1}}$ and $F^l_{\text{2}}$. In this case, $M$ is wavelength-dependent, $M (\lambda) =  (M_{\text{1}} / M_{\text{2}}) \times (\epsilon_{\text{1}} / \epsilon_{\text{2}})$, where $\epsilon_{\text{1}} (\lambda)$ and $\epsilon_{\text{2}} (\lambda)$ represent the transmission factors of the light from images 1 and 2, respectively. $M (\lambda)$ is then estimated at the wavelength of the line.

\begin{figure}[t]
\resizebox{0.95\hsize}{!}{\includegraphics*{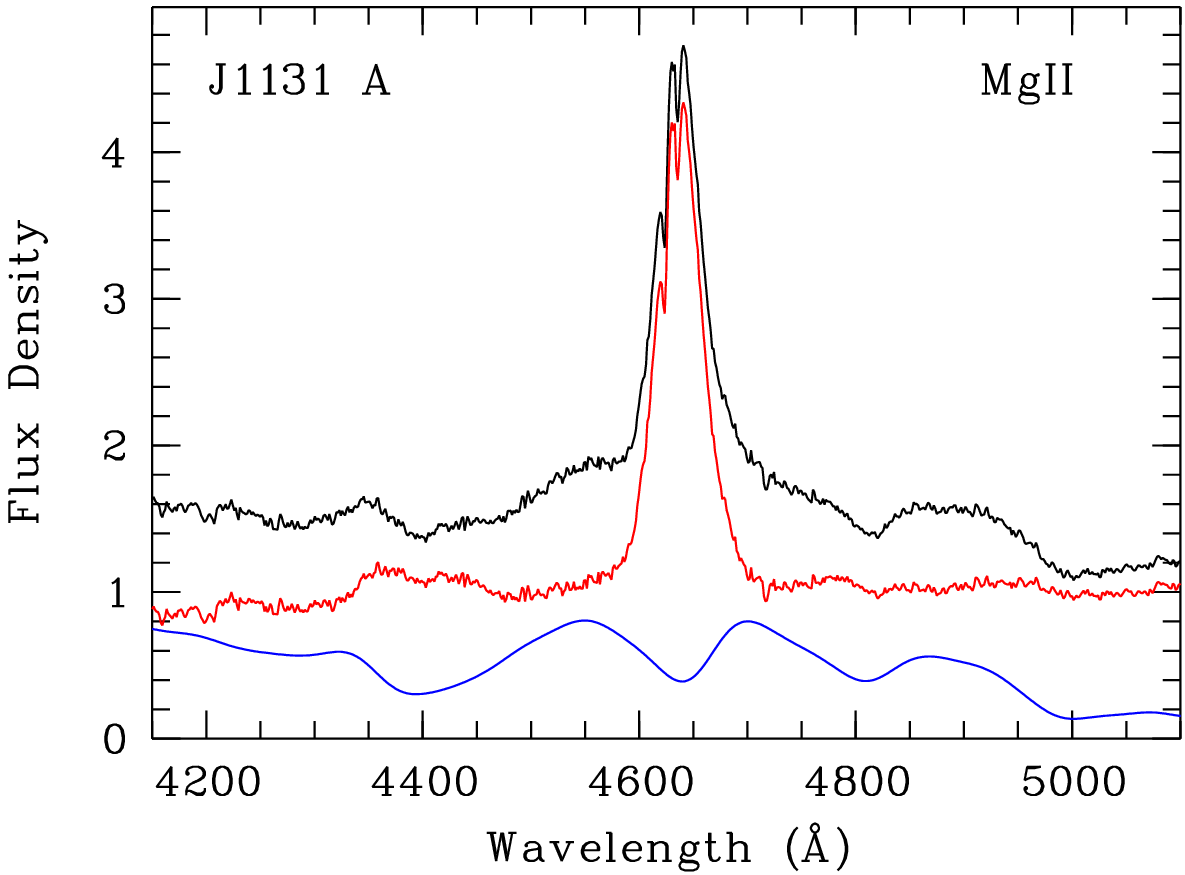}}
\caption{\ion{Fe}{ii}  subtraction from the \ion{Mg}{ii} blend in image A of J1131. The observed spectrum is shown in black, the \ion{Fe}{ii} spectrum in blue, and the spectrum after subtraction in red. The \ion{Fe}{ii} spectrum is the \citet{2006Tsuzuki} template convolved with the BEL width. The flux density is given in arbitrary units.}
\label{fig:fesub}
\end{figure}

The \ion{Mg}{ii} line is usually blended with several \ion{Fe}{ii} emission lines that can form a pseudo-continuum. It is therefore necessary to subtract the \ion{Fe}{ii} spectrum before computing $\mu(v)$ and the indices. In Fig.~\ref{fig:fesub}, we show an example of \ion{Fe}{ii} subtraction. The \ion{Fe}{ii} spectrum is built using the \citet{2006Tsuzuki} template, which is convolved with the BEL full width at half maximum. It is then scaled to remove at best, by eye, the \ion{Fe}{ii} contribution. The same procedure is applied to all targets with an \ion{Mg}{ii} line. As is illustrated in Fig.~\ref{fig:fesub}, the \ion{Fe}{ii} subtraction essentially removes the broadest wings of the \ion{Mg}{ii}+\ion{Fe}{ii} blend. We found that the $\mu(v)$ profile does not depend on a very precise value of the scaling factor, and that it is not very sensitive to the use of another template, in particular the \citet{2001Vestergaard} template. In fact, as long as the high-velocity parts of the line wings are not taken into account, the \ion{Fe}{ii} subtraction has little impact on the $\mu(v)$ profile.

\begin{figure*}[t]
\resizebox{16cm}{!}{\includegraphics*{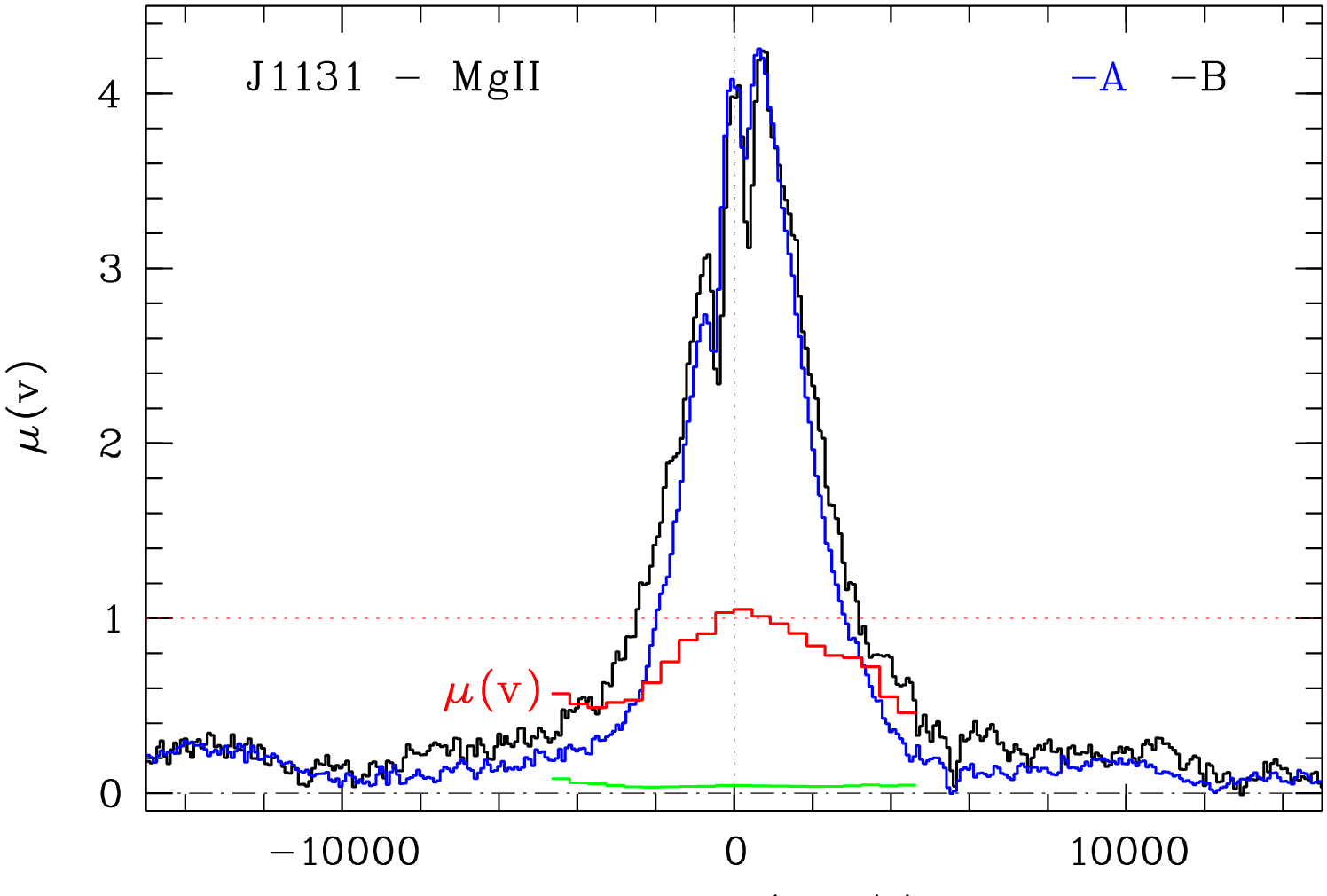}\includegraphics*{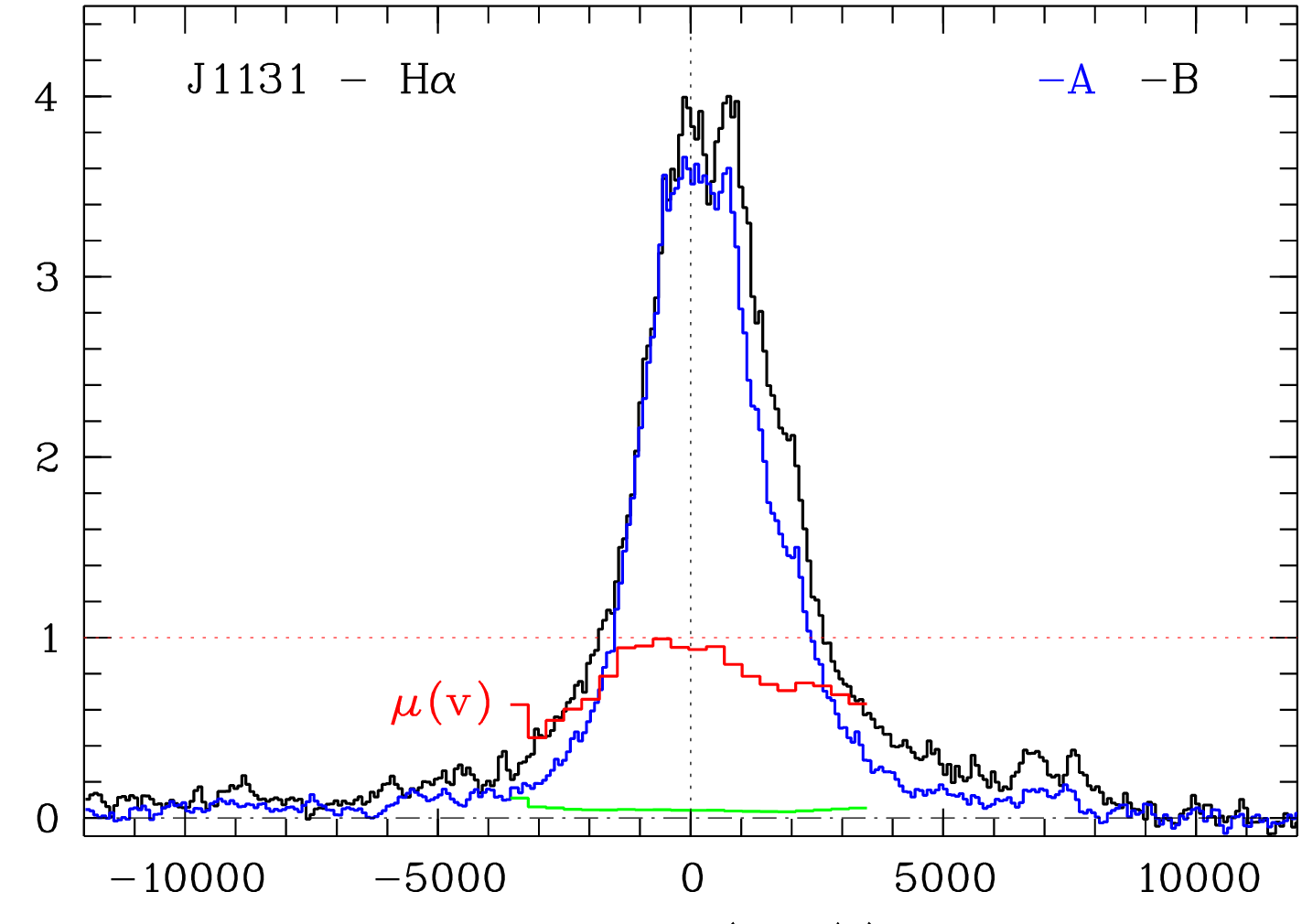}}\\
\resizebox{16cm}{!}{\includegraphics*{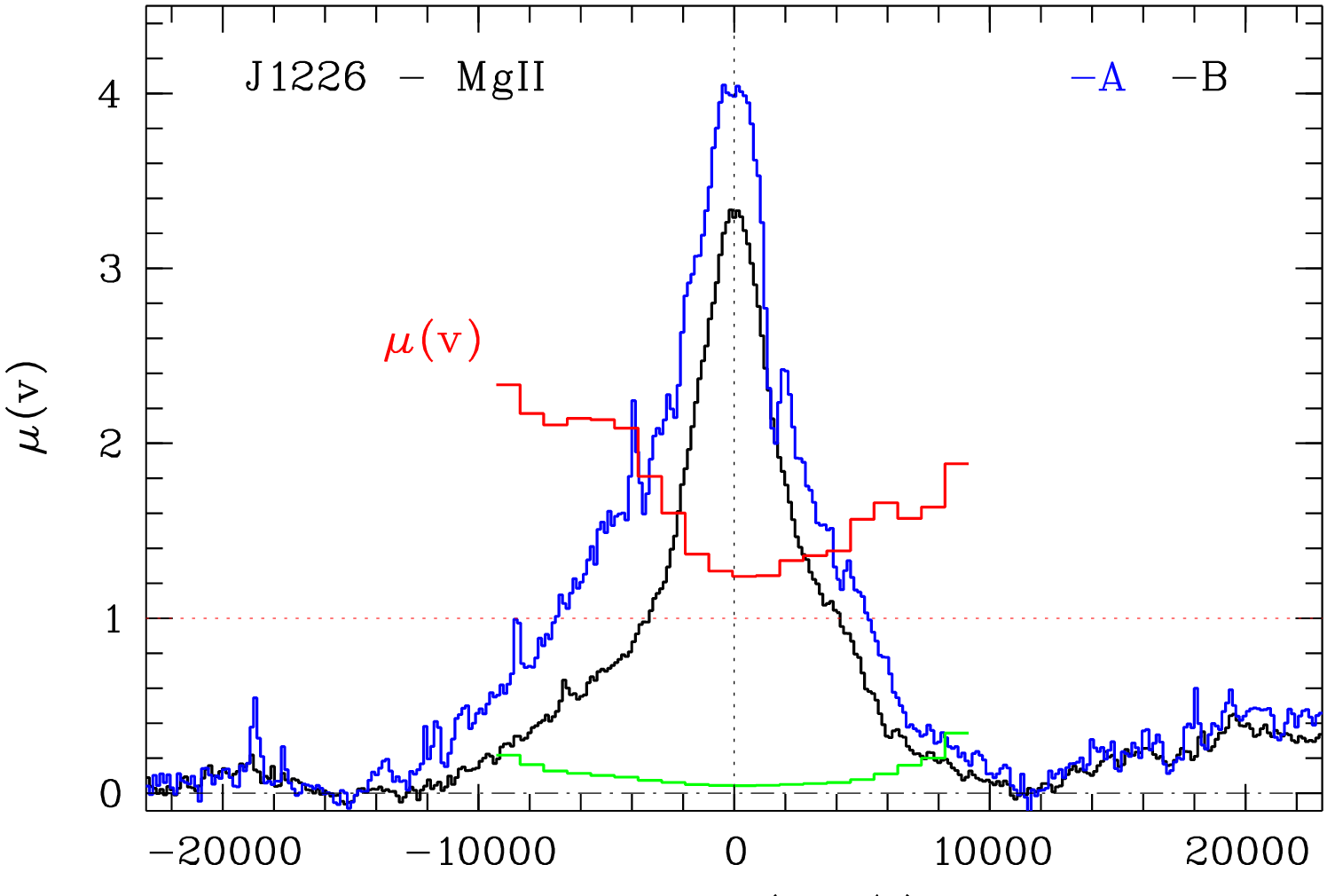}\includegraphics*{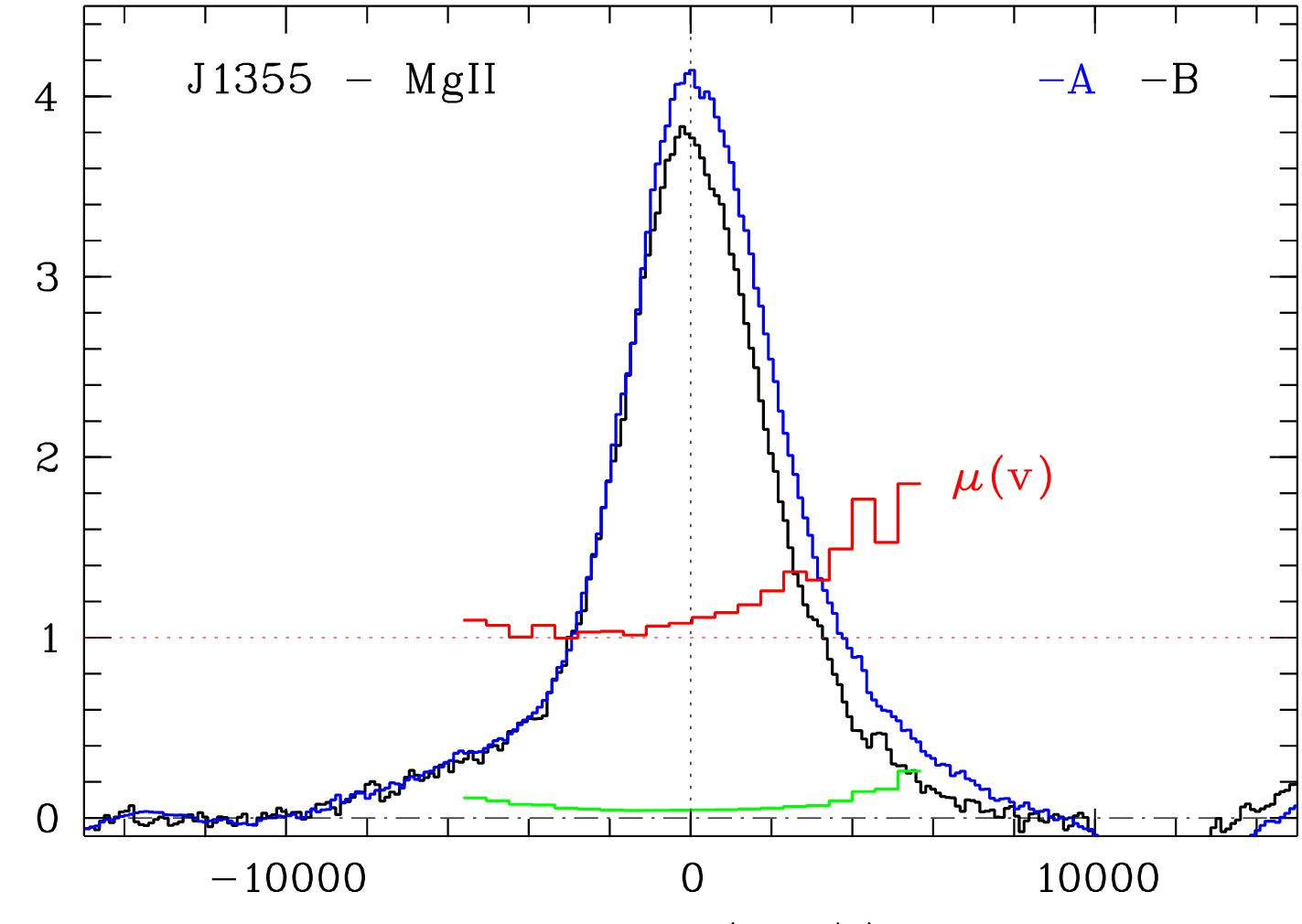}}\\
\resizebox{16cm}{!}{\includegraphics*{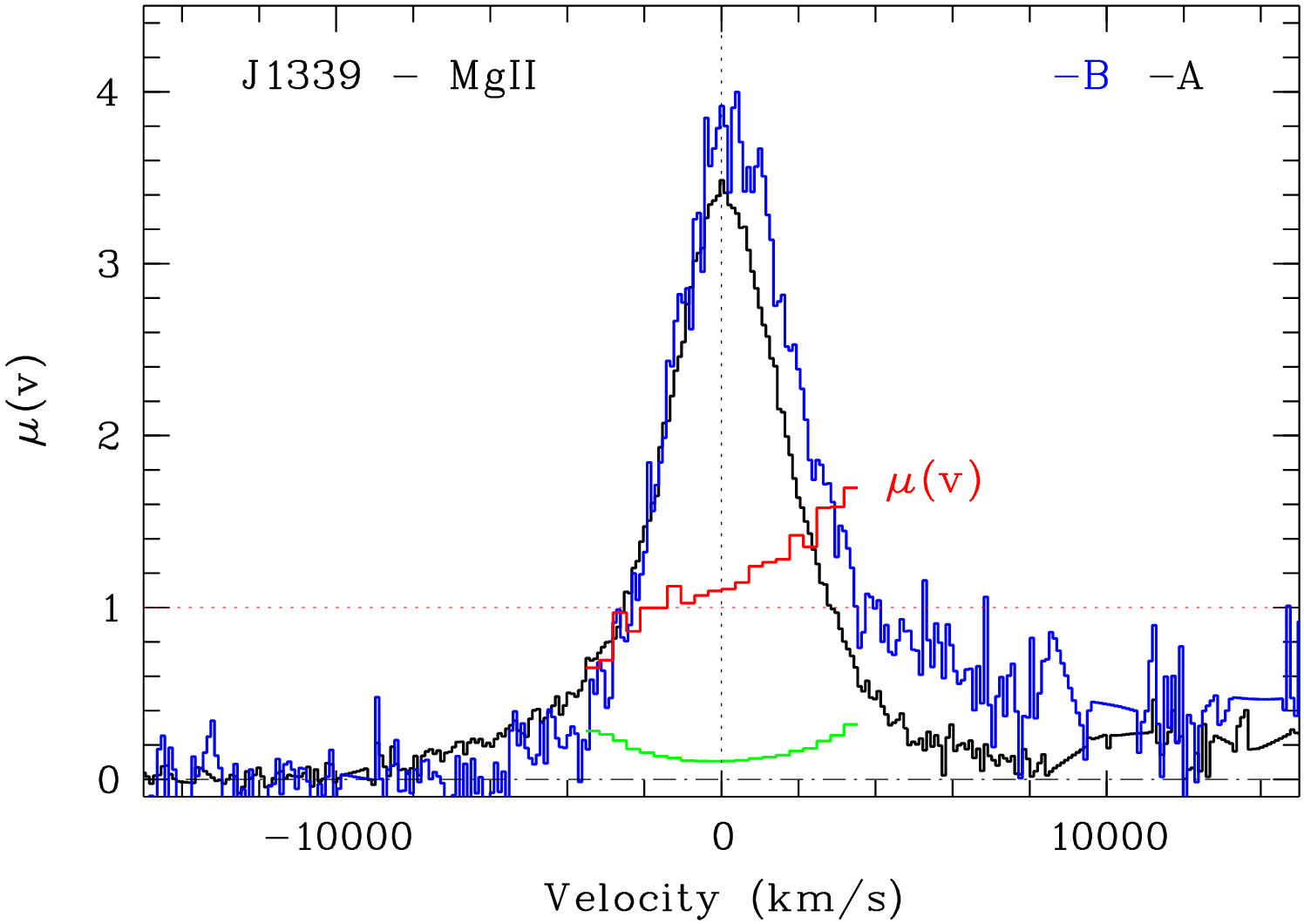}\includegraphics*{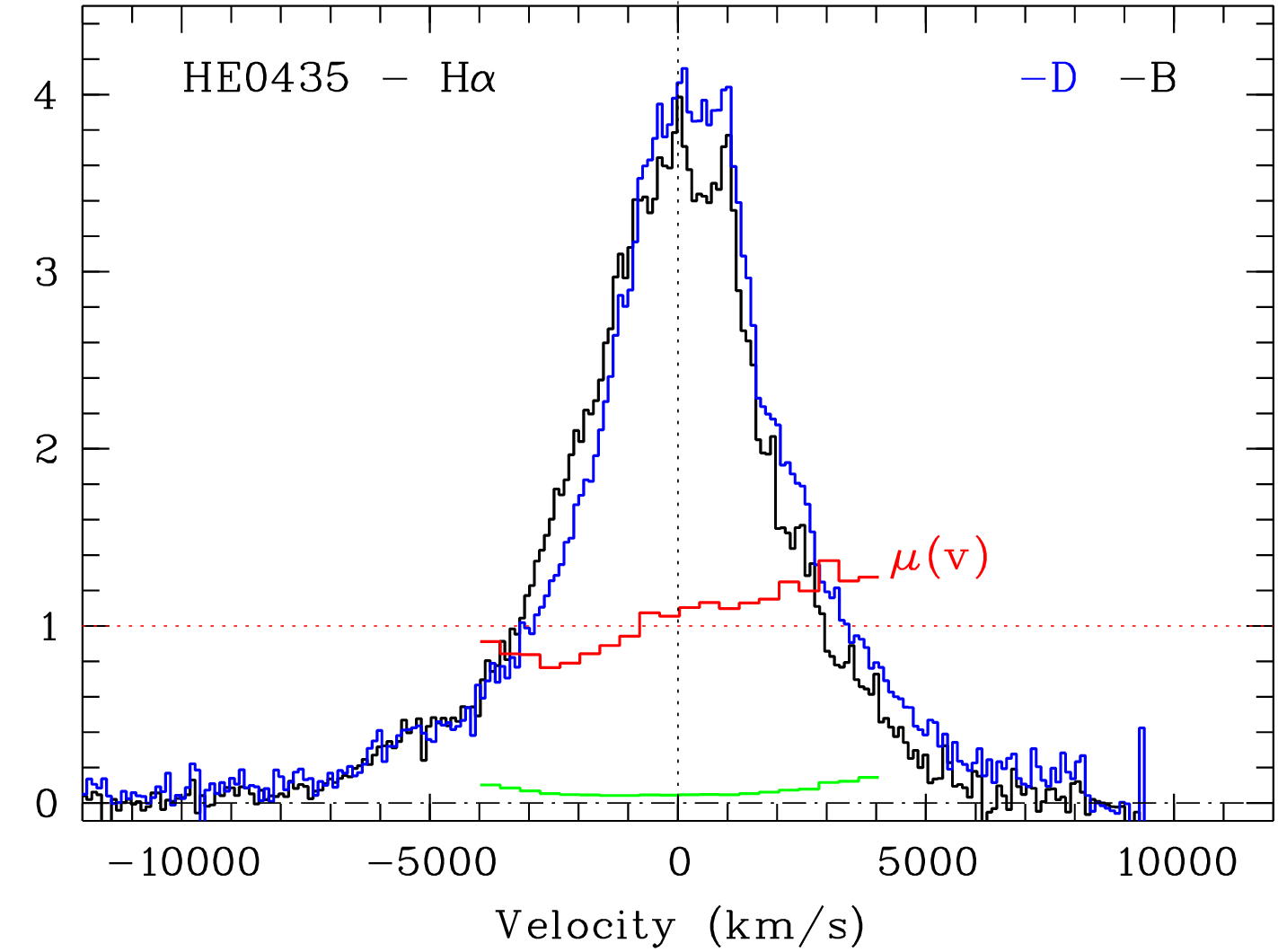}}\\
\caption{$\mu(v)$ magnification profiles of the \ion{Mg}{ii} or H$\alpha$ lines computed from simultaneously recorded spectra of two images of the quasars of our sample. The $\mu(v)$ profiles are shown in red, with the uncertainties in green. The superimposed line profiles from the microlensed image (in blue) and non-microlensed image (in black) were continuum-subtracted, corrected by the $M$ factor, and arbitrarily rescaled. For the \ion{Mg}{ii} line profiles, the \ion{Fe}{ii} spectrum was subtracted using the \citet{2006Tsuzuki} template. The zero-velocity corresponds to the \ion{Mg}{ii} $\lambda$2800  or H$\alpha$  wavelengths at the source redshift.}
\label{fig:muv}
\end{figure*}

To avoid the strong noise in the high-velocity part of $\mu(v)$, where the line flux reaches zero, it is necessary to cut the faintest parts of the line wings.  We considered the parts of the line profiles whose flux density is above $l_{\rm cut} \times F_{\rm peak}$, where $F_{\rm peak}$ is the maximum flux in the line profile and $l_{\rm cut}$ is fixed around 0.1. This cutoff also discards artifacts in the high-velocity wings.  The resulting $\mu(v)$ profiles, binned into 20 spectral elements, are shown in Fig.~\ref{fig:muv} for the different quasars and spectral lines. The measured indices are given in Table~\ref{tab:indices}. In the following, we provide details for each object.

{\em J1131}. \citet{2007Sluse} showed that, in 2003, a clear microlensing effect was present in images A and C, while image B was unaffected. Image C being contaminated by the host galaxy light, we only considered the A, B pair. The time delay between these two images is 1.6$\pm$0.7 day \citep{2020Millon} so that the line distortions can be safely attributed to microlensing. Forbidden lines that originate in the extended narrow-line region are not affected by microlensing and the ratio of their flux measured in different images can be used to estimate the macro-magnification ratio, $M$. Using the [\ion{O}{iii}] $\lambda\lambda 4959,5007$ lines,  \citet{2007Sluse} found that $M_{\text{A}} / M_{\text{B}}$ = 1.97$\pm$0.03, while \citet{2007Sugai} measured  $M_{\text{A}} / M_{\text{B}}$ = 1.63$\pm$0.03. We adopted the mean value, $M_{\text{A}} / M_{\text{B}}$ = 1.80$\pm$0.17, with an uncertainty that encompasses both values. Although the lines are fainter, a similar ratio is found from [\ion{Ne}{v}] $\lambda\lambda 3346,3426$: $M_{\text{A}} / M_{\text{B}} \simeq$ 1.85. This indicates that there is no significant wavelength-dependent differential extinction between images A and B, as was suggested by \citet{2007Sluse}. We then used $M(\text{\ion{Mg}{ii}})$ = $M(\text{H}\alpha)$ = 1.80$\pm$0.17 to compute  $\mu(v)$ and the indices for both the \ion{Mg}{ii} and H$\alpha$ lines.
The underlying continuum was measured in two windows on each side of the line profiles, $[-33,-28]$ and $[+23,+27]$ 10$^3$~km~s$^{-1}$ for \ion{Mg}{ii} and  $[-20,-17]$ and $[+9,+12]$ 10$^3$~km~s$^{-1}$ for H$\alpha$, interpolated by a straight line, and subtracted from the line profiles.  For both lines, we cut the faintest parts of the line profile using $l_{\rm cut}$ = 0.1.

{\em J1226}. The time delay between images A and B is around 30 days \citep{2020Millon,2021Donnan}, which is short enough to suggest that the line profile distortions observed in the \ion{Mg}{ii} line are due to microlensing. A priori, either A or B can be microlensed. However, the analysis of the microlensing  magnification maps computed for images A and B (Sect.~\ref{sec:models}) shows that high magnifications can be generated from the image A maps and reproduce the observations, while the equivalent strong demagnifications that would be needed for image B cannot be reached.  We therefore assume that A is the microlensed image. The macro-magnification factor, $M_{\text{A}} / M_{\text{B}}$ = 1.14$\pm$0.10, was evaluated from the ratio of the [\ion{O}{ii}] $\lambda 3727$ lines. It is in good agreement with  $M_{\text{A}} / M_{\text{B}}$ = 1.03$\pm$0.18 measured at radio wavelengths \citep{2024Jackson}, suggesting the absence of significant wavelength-dependent differential extinction between images A and B. We then used $M(\text{\ion{Mg}{ii}})$ = 1.14$\pm$0.10 to compute $\mu(v)$ and the indices.
The continuum windows on each side of the \ion{Mg}{ii} line profile were $[-17,-15]$ and $[+11,+12]$ 10$^3$~km~s$^{-1}$.  The faintest parts of the line profile were cut using $l_{\rm cut}$ = 0.05. 

{\em J1355}. The distortions of the \ion{Mg}{ii} line profile observed in 2005 \citep{2012Sluse} were still observed in 2008 \citep{2020Rojas}, and hence over a period of time much longer than the time delay between images A and B, which is around 60 days \citep{2020Millon,2021Donnan}. This confirms the microlensing interpretation, and either A or B can be microlensed. As there is no information on which image is microlensed, we have assumed that image A is microlensed (see Sect.~\ref{sec:radlum} for further discussion). Using the macro-micro decomposition (MmD) method, \citet{2012Sluse} measured $M_{\text{A}} / M_{\text{B}}$ = 2.95$\pm$0.25 at the wavelength of \ion{Mg}{ii}.  The MmD provides a direct measurement of $M$ that contains both the true macro-magnification ratio and the differential extinction. Since the value derived for \ion{Mg}{ii} was found to be in excellent agreement with the [\ion{Ne}{v}] $\lambda\lambda 3346,3426$ line ratio measured from images A and B \citep{2012Sluse}, we adopted $M(\text{\ion{Mg}{ii}})$ = 2.95$\pm$0.25 to compute $\mu(v)$ and the indices.
The continuum windows on each side of the \ion{Mg}{ii} line profile were $[-14,-10]$ and $[+8,+10]$ 10$^3$~km~s$^{-1}$, avoiding the atmospheric absorption at $11 \times 10^3$~km~s$^{-1}$.  The faintest parts of the line profile were cut using $l_{\rm cut}$ = 0.075.

{\em J1339}. Image B is strongly affected by microlensing, with clear line profile distortions with respect to image~A \citep{2016Goicoechea,2021Shalyapin}. The distortions are observed over a period of at least a few years, much longer than the time delay of 47 days between the two images \citep{2021Shalyapin}, which supports the microlensing interpretation. The determination of the macro-magnification factor, $M_{\text{B}} / M_{\text{A}}$ =  0.20$\pm$0.03, from the [\ion{O}{iii}] line ratio is discussed in \citet{2024Hutsemekers}. In this object, the extinction of image A must be taken into account \citep{2016Goicoechea,2021Shalyapin}. Interpolating $\epsilon_{\text{A}}$ at the wavelength of \ion{Mg}{ii}, as was done for \ion{C}{iv} \citep{2024Hutsemekers}, we derived $M(\text{\ion{Mg}{ii}}) = 0.26\pm0.04$, which was used in the computation of $\mu(v)$ and the indices. The $\mu(v)$ profile was taken as rather narrow in velocity to avoid the strong noise that affects the red wing of the \ion{Mg}{ii} line.
The continuum windows on each side of the \ion{Mg}{ii} line profile were $[-15.5,-13.5]$ and $[+25.0,+27.5]$ 10$^3$~km~s$^{-1}$. The faintest parts of the line profile were cut using $l_{\rm cut}$ = 0.17.

{\em HE0435}. In this quadruply imaged quasar, image D was found microlensed and image B unaffected by microlensing \citep{2014Braibant}. The short delay between these two images of about 5 days \citep{2020Millon,2021Donnan} supports the microlensing interpretation. The microlensing analysis previously reported in \citet{2019Hutsemekers} is revised here, based on the updated method (the use of $\mu(v)$ and not only the four indices), and a different value of $M(\text{H}\alpha)$. In \citet{2014Braibant}, $M(\text{H}\alpha)$ = 0.47$\pm$0.03 was determined with the MmD method, and used in \citet{2019Hutsemekers}. However, this method can give inaccurate values if parts of the line profile are magnified and other parts are demagnified \citep{2010Hutsemekers,2024Hutsemekers}. After the \citet{2014Braibant} paper, \citet{2017Nierenberg} measured $M_{\text{D}} / M_{\text{B}}$ =  0.67$\pm$0.05 from the [\ion{O}{iii}] line ratio, in excellent agreement with $M_{\text{D}} / M_{\text{B}}$ =  0.61$\pm$0.09 derived from radio observations \citep{2015Jackson}. This agreement indicates the absence of significant differential extinction at the wavelength of H$\alpha$. We then adopted $M(\text{H}\alpha)$ = 0.67$\pm$0.05. With this value, $\mu(v)$ shows magnification of the red part of the line profile and demagnification of the blue part.
The continuum windows on each side of the H$\alpha$ line profile were $[-20,-17]$ and $[+8,+9]$ 10$^3$~km~s$^{-1}$.  The faintest parts of the line profile were cut using $l_{\rm cut}$ = 0.16.

The $\mu(v)$ magnification profiles shown in Fig.~\ref{fig:muv} have an S/N between 5 (line wings) and 25 (line core), except for J1339 where the S/N is a factor of two lower. Together with those obtained in Q2237$+$0305, J1004$+$4112, and J1138$+$0314 \citep{2021Hutsemekers,2023Hutsemekers,2024Hutsemekers}, the $\mu(v)$ magnification profiles show a large diversity. While the line core is often less magnified, the blue and red wings can either both be magnified, both be demagnified, or one can be magnified and the other one demagnified, or they can simply be unaffected. In a given object, the $\mu(v)$ profiles of different lines have roughly similar shapes, although the high-ionization \ion{C}{iv} profiles show stronger magnifications than the low-ionization \ion{Mg}{ii} and H$\alpha$ lines (cf. J1339 and Q2237$+$0305).

\section{Microlensing simulations}
\label{sec:models}

We computed the effect of gravitational microlensing on the BEL profiles by convolving, in the source plane, the emission from BLR models with microlensing magnification maps. The microlensing simulations and the comparison to observations were carried out in the manner described in \cite{2023Hutsemekers,2024Hutsemekers}. The method is essentially based on \cite{2017Braibant}, in which the models are detailed, and \cite{2019Hutsemekers}, in which the probabilistic analysis was developed.

For the BLR models, we considered a rotating Keplerian disk (KD), a biconical, radially accelerated polar wind (PW), and a radially accelerated equatorial wind (EW),  with inclinations with respect to the line of sight of $i$ = 22\degr, 34\degr, 44\degr, and 62\degr. Using the radiative transfer code STOKES \citep{2007Goosmann,2012Marin,2014Goosmann}, we generated 20 BLR monochromatic images, which correspond to 20 spectral bins in the line profile. The BLR models have an emissivity of $\varepsilon = \varepsilon_0 \, (r_{\text{in}}/r)^q$, where $r_{\text{in}}$ is the BLR inner radius of the model, and $q=3$ or $q=1.5$. We considered a range of $r_{\text{in}}$ expressed in terms of the microlensing Einstein radius in the source plane, $r_{\text E}$.  For a lens of mass $\mathcal{M}$,
\begin{equation}
  \label{eq:re}
  r_{\text E} = \sqrt{4 \; \frac{G\mathcal{M}}{c^2} \frac{D_S D_{LS}}{D_L}} \, ,
\end{equation}
where $D_S$, $D_L$, and $D_{LS}$ are the source, lens, and lens-source angular diameter distances, respectively. We used between 14 and 18 values of $r_{\text{in}}$, from 0.01 to 4 $r_{\text E}$ depending of the object. In all cases, the outer radius of the BLR was fixed to $r_{\text{out}} = 10 \, r_{\text{in}}$. The continuum source was modeled as a disk of constant surface brightness (uniform disk), with the same inclination as the BLR, and  with outer radii ranging from $r_s =$ 0.02 to 5~$r_{\text E}$.

\begin{table}[b]
\caption{Macro-model parameters of the lensed images.}
\label{tab:kappagamma}
\renewcommand{\arraystretch}{1.2}
\centering
\begin{tabular}{lcccc}
\hline\hline
Image  & $\kappa$ & $\gamma$ & $\kappa_{\star}/\kappa$ & Ref.  \\
\hline
HE0435D            & 0.59 & 0.64 & 0.07, 0.21       & (1, 2) \\
J1226A             & 0.67 & 0.61 & 0.07, 0.21 & (1, 2)\\
J1355A             & 0.35 & 0.39 & 0.07, 0.21 & (1, 2) \\
J1339B$^\dagger$    & 0.86 & 0.45 & 0.11       & (3) \\ 
J1339B$^\dagger$    & 0.63 & 0.90 & 0.52       & (3) \\
J1131A             & 0.47 & 0.57 & 0.07, 0.3  & (4, 2, 5) \\  
\hline
\end{tabular}
\tablefoot{Convergence, $\kappa$,  shear, $\gamma$, and local stellar fraction, $\kappa_{\star}/\kappa$, are the model parameters. The references associated with the macro-model parameters and the local stellar fraction (when different) are indicated in the last column. (1)  \cite{2012Sluse},  (2)  \cite{2015Jimeneza}, (3)  \cite{2021Shalyapin},  (4)  \cite{2012aSluse}, (5)  \cite{2010Dai}. $\dagger$:~Two plausible sets of models have been considered corresponding to two sets of ($\kappa$, $\gamma$).}
\end{table}

Each microlensing simulation is based on a macro-model of the lensing galaxy that reproduces the main strong lensing observables of each system. This macro-model enabled us to predict the convergence, $\kappa$ (i.e., the normalized mass surface density), and shear, $\gamma$, at the positions of the lensed images. The fraction of the local convergence in the form of stars, $\kappa_{\star} / \kappa$, is based on realistic priors. For each lensed image, we thus have a set of three parameters ($\kappa, \gamma, \kappa_{\star}$), which we used to compute the magnification maps. This last step was performed using the ray-tracing code developed by \citet{1999Wambsganss}. The parameters of the lensed images are given in Table~\ref{tab:kappagamma}. For all systems but J1339, the macro-model consists of an isothermal model+external shear. For J1339, \citet{2021Shalyapin} presented an ensemble of ten lens models constituted of the sum of a stellar and a dark matter component embedded into an external shear field. These ten models overfit the data but self-consistently predict the stellar fraction, $\kappa_{\star}/\kappa$,  at the image positions. We elected two models that bound the expected fraction of total stellar mass in elliptical galaxies \citep[e.g.,][]{2022Shajib}. Specifically, we chose a model with a stellar fraction of $\kappa_{\star}/\kappa= 0.11$ and a model with $\kappa_{\star}/\kappa= 0.52$. For the other systems, we simply used $\kappa_{\star}/\kappa$ = 0.07 and 0.21. These bounds match the constraints on the mean local stellar mass fraction derived from X-ray microlensing studies of lensed quasars \citep{2012Pooley, 2015Jimeneza}. For J1131, we chose $\kappa_{\star}/\kappa=0.3$ as an upper bound based on the analysis of the microlensing in this system by \citet{2010Dai}. As is explained hereafter, our results depend weakly on $\kappa_{\star} / \kappa$, provided that it remains within the selected bounds. 

To investigate the impact of the orientation of the symmetry axis of the BLR models relative to the caustic network, the maps were rotated clockwise by $\theta$ = 15\degr, 30\degr, 45\degr, 60\degr, 75\degr\  , and 90\degr\ with respect to the BLR model axis. $\theta$ = 0\degr\ corresponds to the caustic elongation and shear direction perpendicular to the BLR model axis. The final maps extend over a $100 \, r_{\text E} \times 100 \, r_{\text E}$ area of the source plane and are sampled by $10000 \times 10000$ pixels.

Distorted line profiles were computed by convolving, for a given BLR size, the magnification maps with the monochromatic (isovelocity) images of the BLR. Simulated $\mu(v)$ profiles were then obtained for each position of the BLR on the magnification maps. The continuum-emitting region was similarly convolved.  The likelihood that the simulations reproduce the observables, $\mu^{cont}$, and the 20 spectral elements of $\mu(v)$, was finally computed for each set of parameters.

\section{Results}
\label{sec:results}

\begin{figure*}[t]
\resizebox{\hsize}{!}{\includegraphics*{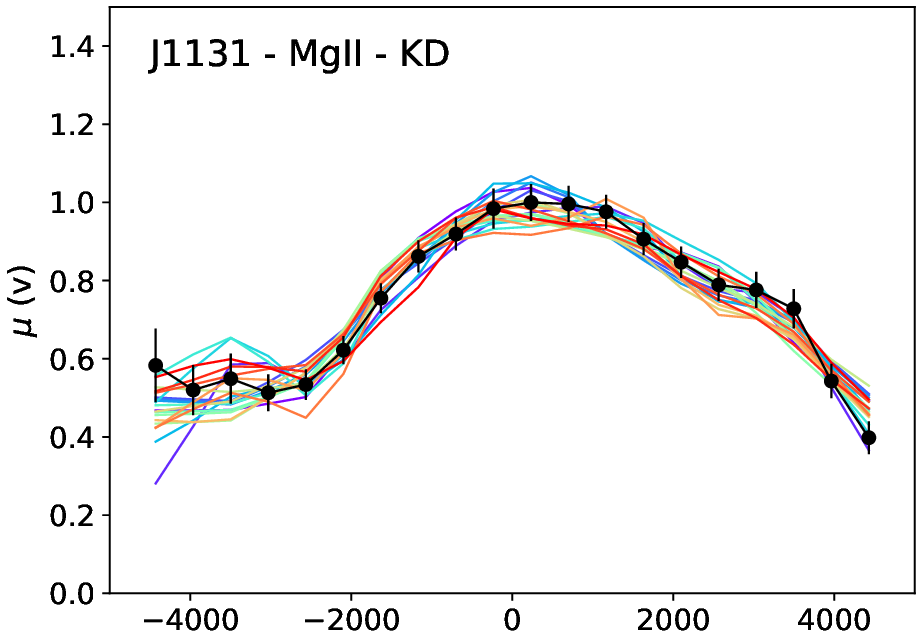}\includegraphics*{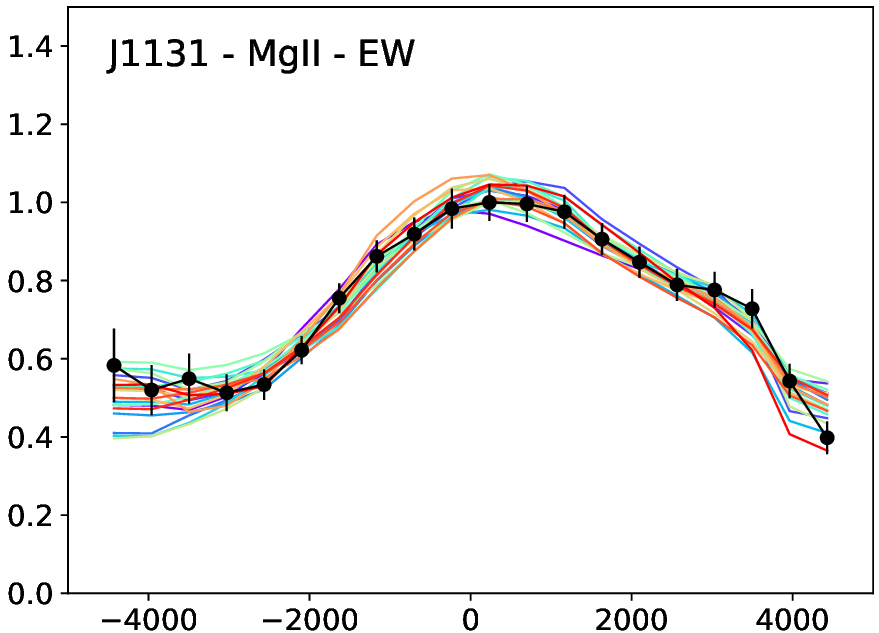}\includegraphics*{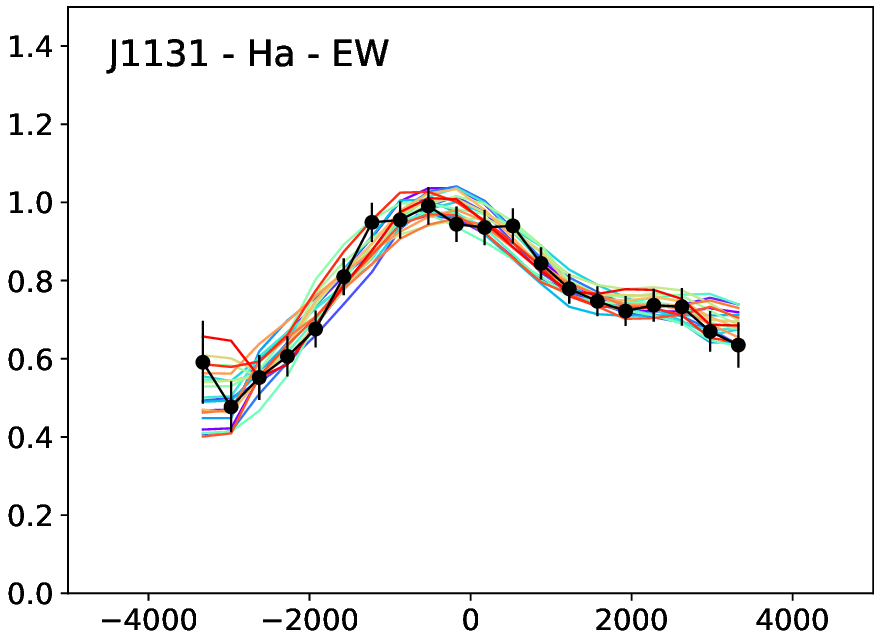}}\\
\resizebox{\hsize}{!}{\includegraphics*{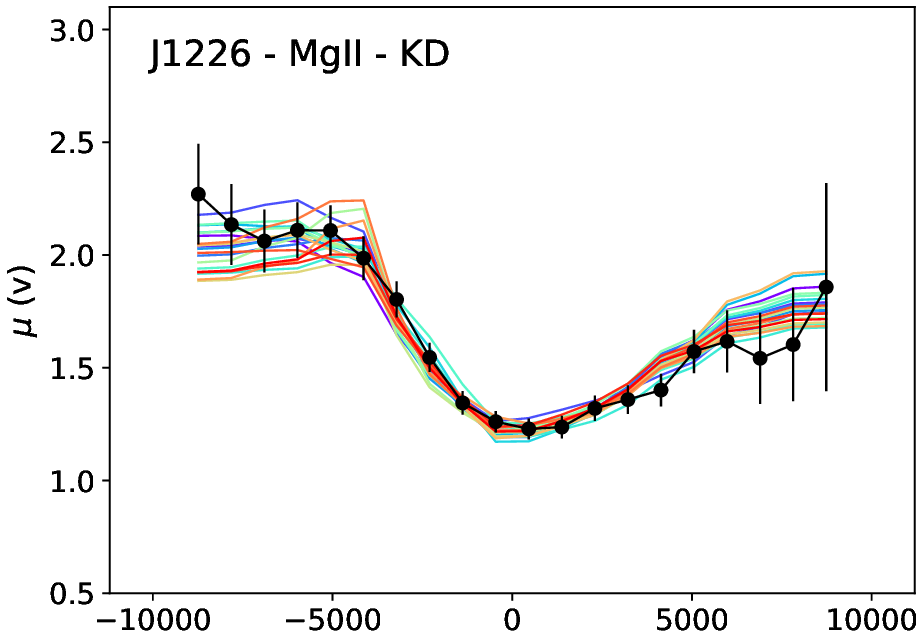}\includegraphics*{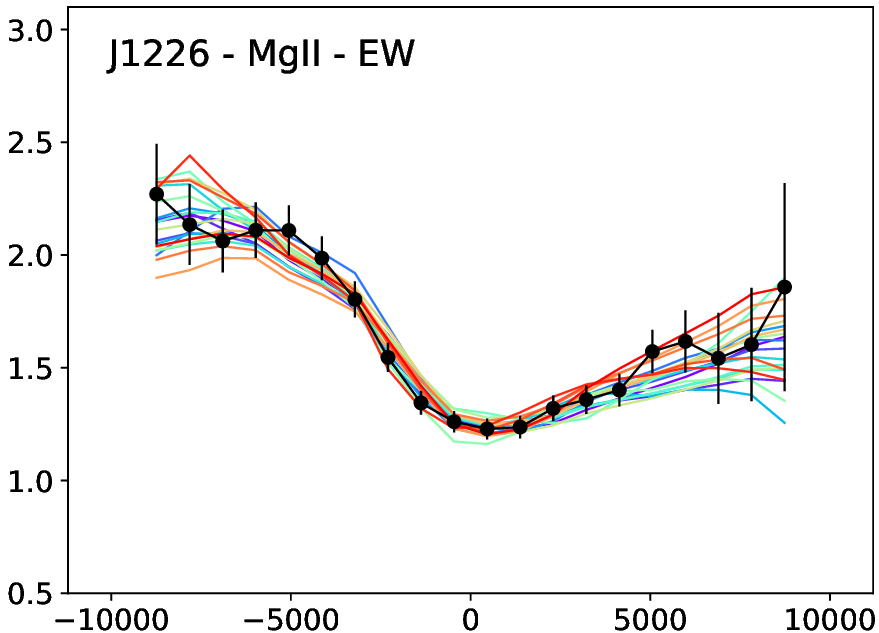}\includegraphics*{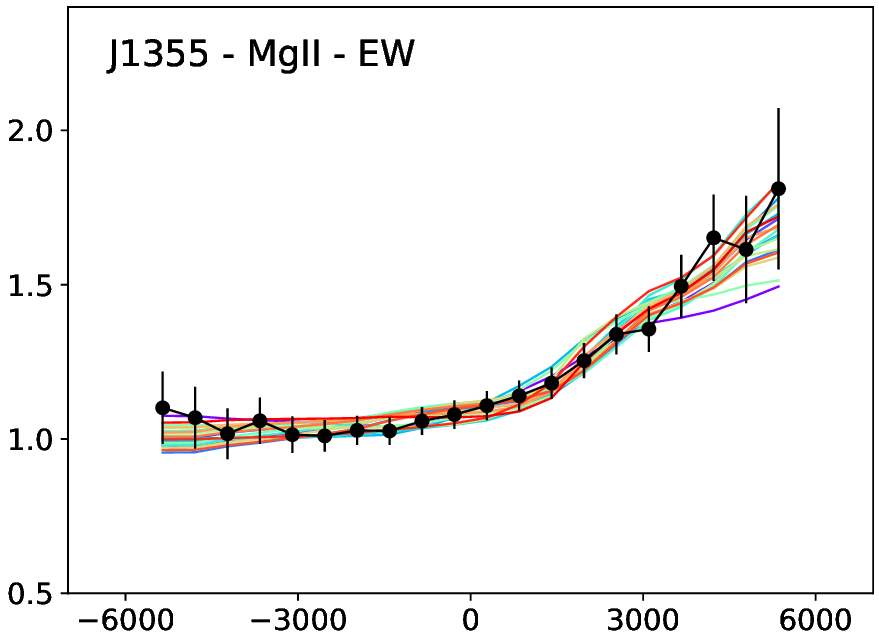}}\\
\resizebox{\hsize}{!}{\includegraphics*{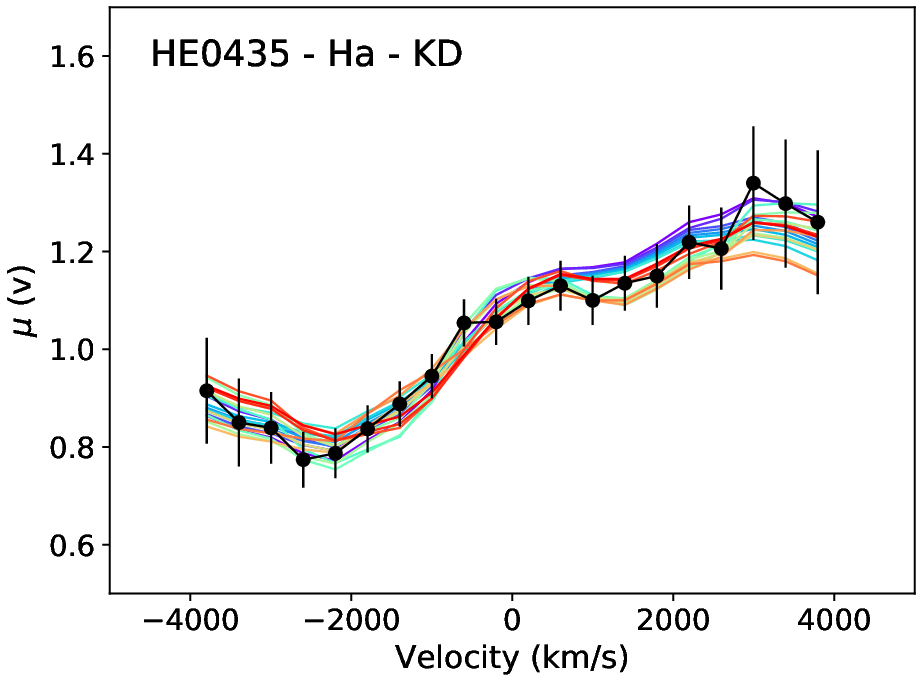}\includegraphics*{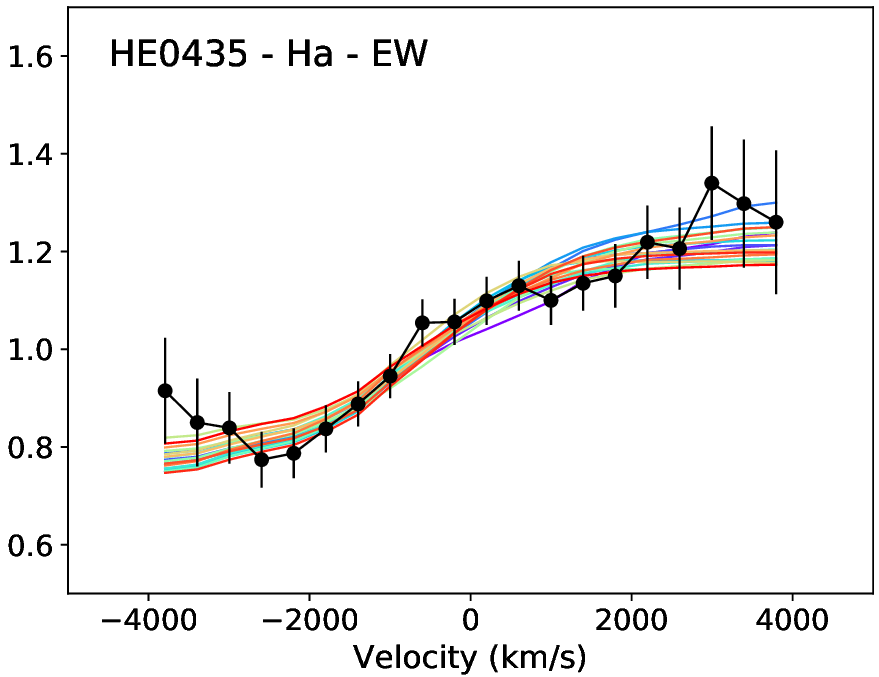}\includegraphics*{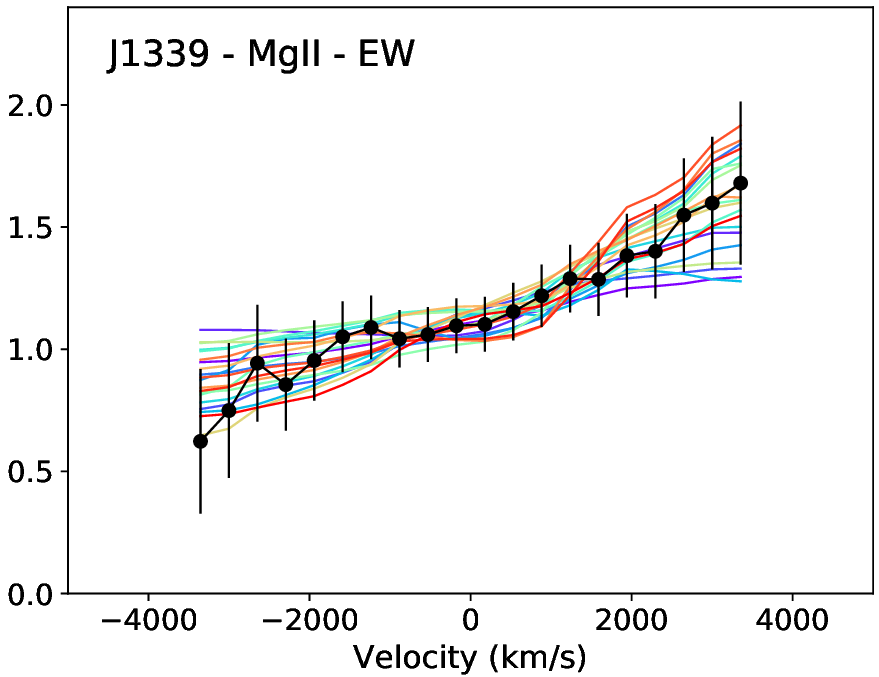}}
\caption{Examples of 20 simulated $\mu(v)$ profiles (in color) that fit the $\mu(v)$ magnification profiles (in black) of the \ion{Mg}{ii} or H$\alpha$ emission lines observed in J1131, J1226, J1355, J1339, and HE0435. The observed $\mu(v)$ profiles can be slightly different from those shown in Fig.~\ref{fig:muv} due to their resampling on the simulated profile velocity grid. The models are either EW or KD. The magnification maps with the smallest $\kappa_{\star} / \kappa $ value were used. The maps are oriented at $\theta$ = 60\degr\ for EW models and $\theta$ = 0\degr\ for KD models, except for HE0435, where $\theta$ = 60\degr\ for KD and $\theta$ = 0\degr\ for EW (Table~\ref{tab:proba}). In all models,  $i = 34\degr$, $q = 1.5$, and $r_{\rm in} =  0.1 \, r_{\rm E}$. }
\label{fig:fitmuv}
\end{figure*}

\begin{table}[]
\caption{Probabilities of the \ion{Mg}{ii} and H$\alpha$ BLR models}
\label{tab:proba}
\renewcommand{\arraystretch}{1.1}
\centering
\begin{tabular}{lccccccc}
\hline\hline
     & \multicolumn{3}{c}{$\theta \leq 30\degr$} &  & \multicolumn{3}{c}{$\theta \geq 60\degr$} \\
\hline
          & KD & PW & EW &  & KD & PW & EW  \\
\hline
     \multicolumn{8}{c}{J1131 $-$  \ion{Mg}{ii}} \\
\hline
          22\degr         & 35 &  0 & 0  &  & 0 & 0 &   7    \\
          34\degr         & 27 &  0 & 0  & & 0 & 0 &  29    \\
          44\degr         & 10 &  7 & 0  &  & 0 & 0 &  52    \\
          62\degr         & 17 &  5 & 0  &  & 0 & 0 &  12    \\
          All $i$         & 89 & 11 & 0  &  & 0 & 0 & 100    \\
\hline
     \multicolumn{8}{c}{J1131 $-$  H$\alpha$} \\
\hline
          22\degr         & 48 &  0 & 0  &  & 0 & 0 &   2    \\
          34\degr         & 15 &  1 & 0  &  & 0 & 0 &  26    \\
          44\degr         &  9 &  8 & 0  &  & 0 & 0 &  50    \\
          62\degr         & 11 &  8 & 0  &  & 0 & 0 &  21    \\
          All $i$         & 83 & 16 & 0  &  & 0 & 0 & 100    \\
\hline
     \multicolumn{8}{c}{J1226 $-$  \ion{Mg}{ii}} \\
\hline
          22\degr         & 48 & 0 & 0  &  & 1 & 0 & 4     \\
          34\degr         & 19 & 3 & 0  &  & 0 & 0 & 61     \\
          44\degr         & 15 & 3 & 0  &  & 0 & 1 & 30     \\
          62\degr         & 11 & 0 & 1  &  & 0 & 0 & 3     \\
          All $i$         & 94 & 6 & 1  &  & 1 & 1 & 98     \\
\hline
     \multicolumn{8}{c}{J1355 $-$  \ion{Mg}{ii}} \\
\hline
          22\degr         & 8 & 0 & 15  &  & 2 & 0 & 43     \\
          34\degr         & 4 & 0 & 17  &  & 1 & 0 & 28     \\
          44\degr         & 3 & 0 & 18  &  & 0 & 0 & 18     \\
          62\degr         & 4 & 8 & 23  &  & 0 & 1 &  7    \\
          All $i$        & 20 & 8 & 72  &  & 3 & 1 & 95    \\
\hline
     \multicolumn{8}{c}{J1339 $-$  \ion{Mg}{ii}} \\
\hline
          22\degr         & 6 & 0 & 19  &  & 8 & 0 & 31     \\
          34\degr         & 3 & 1 & 13  &  & 4 & 1 & 19     \\
          44\degr         & 3 & 3 & 12  &  & 3 & 2 & 15     \\
          62\degr         & 5 & 17 & 16 &  & 2 & 5 & 11     \\
          All $i$         & 17 & 22  & 61  &  & 16  & 8 & 76     \\
\hline
     \multicolumn{8}{c}{HE0435 $-$   H$\alpha$} \\
\hline
          22\degr         & 4 & 0 & 9   &  & 44 & 0 & 3      \\
          34\degr         & 2 & 0 & 18  &  & 24 & 0 & 9     \\
          44\degr         & 1 & 0 & 19  &  & 11 & 0 & 4     \\
          62\degr         & 2 & 30 & 15 &  &  4 & 0 & 1    \\
          All $i$         & 9 & 30 & 62 &  & 83 & 0 & 17     \\
\hline
\end{tabular}
\tablefoot{The probabilities are given in percent. $\theta$ is the angle between the BLR axis and the magnification map orientation.}
\end{table}

Examples of simulated $\mu(v)$ profiles that fit the observed $\mu(v)$ profiles are shown in Fig.~\ref{fig:fitmuv}. As has already been noticed for the \ion{C}{iv} emission line \citep{2023Hutsemekers,2024Savic,2024Hutsemekers}, $\mu(v)$ profiles of various shapes can be reproduced by the microlensing-induced distortions of line profiles generated in the framework of the simple KD and EW models considered here. We stress that, in several cases, the observed line profile distortions can be simulated using either the KD or EW models, depending on the orientation of the magnification map with respect to the BLR model axis. As was first discussed in the case of J1004$+$4112  \citep{2023Hutsemekers}, this shows that the $\mu(v)$ microlensing magnification profiles strongly depend on the position of the isovelocity parts of a BLR model with respect to the caustic network, and not only on their different effective sizes, as is often assumed (for a more detailed discussion of the latter point, see Appendix~\ref{sec:notonlysize}).

\subsection{Most likely broad-line region models}
\label{sec:proba}

The relative probability of the different models was obtained by marginalizing the likelihood over all parameters except the model and its inclination (for details, see \citealt{2019Hutsemekers,2023Hutsemekers}). The probabilities are given in Table~\ref{tab:proba} for $\theta \leq 30\degr$ and $\theta \geq 60\degr$ separately. Since the preferred models are found to be essentially independent of the fraction of compact matter, the probabilities computed with the different maps were merged.

For J1131, the most likely model for both the \ion{Mg}{ii} and H$\alpha$ BLRs is either KD or EW, depending on the map orientation relative to the BLR axis. For J1226, we see the same dependence of the most likely model on the map orientation. For the \ion{Mg}{ii} BLR in J1355 and J1339, the EW model is preferred whatever the map orientation. The same result was obtained for the J1339 \ion{C}{iv} BLR \citep{2024Hutsemekers}. We note that for the quasar Q2237+0305, the KD model is the most likely, whatever the orientation of the map \citep{2024Savic}. For the  H$\alpha$ BLR in HE0435, either KD or EW dominates, depending on the map orientation. For all objects, the PW model has a lower probability. The inclination, $i$, is poorly constrained, but it is smaller than 45\degr , as is expected for type~1 quasars. As for the high-ionization \ion{C}{iv} BLR, we can conclude that disk geometries with kinematics dominated by either Keplerian rotation or equatorial outflow best reproduce the microlensing effects on the low-ionization \ion{Mg}{ii} and H$\alpha$ emission line profiles. We emphasize that the KD and EW models could be discriminated if we had knowledge of the orientation of the BLR axis on the sky, such as the one given by the orientation of the radio jet.

\begin{figure*}[p]
\resizebox{17cm}{!}{\includegraphics*{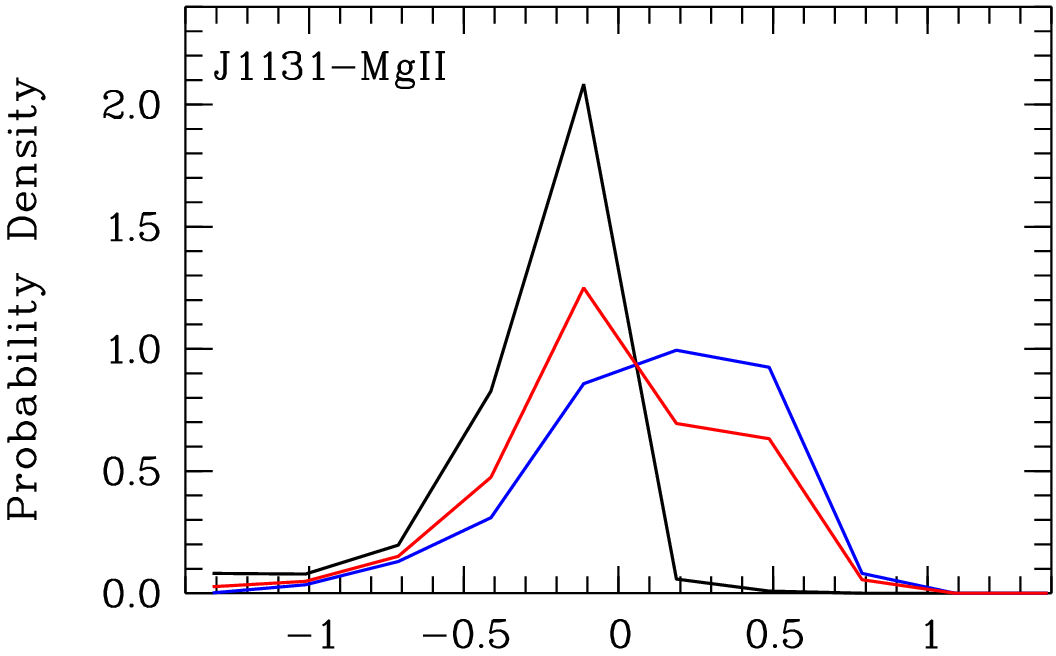}\includegraphics*{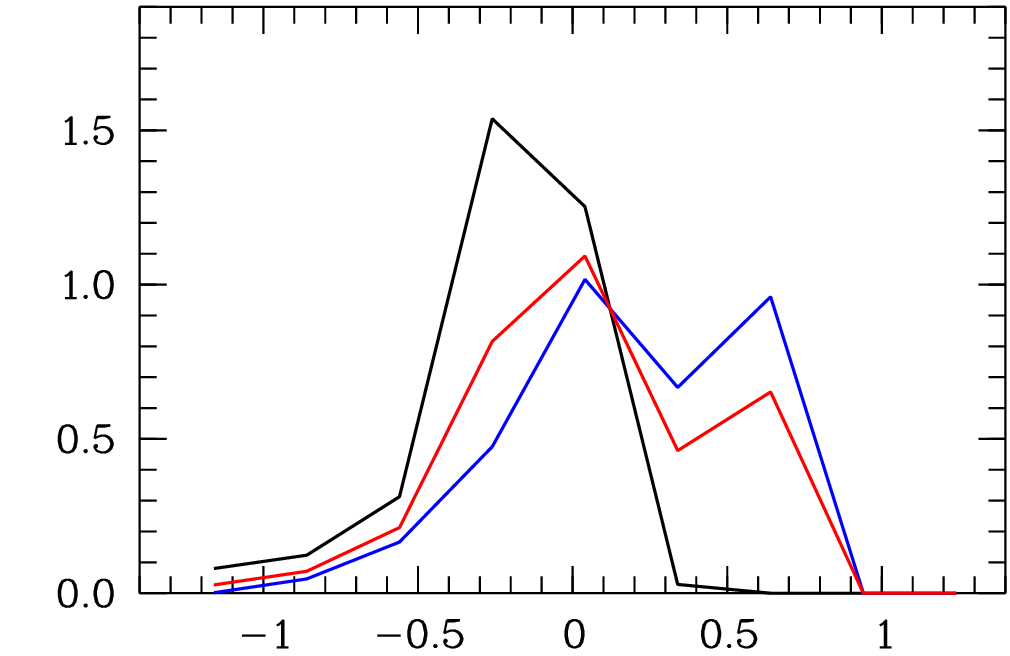}\includegraphics*{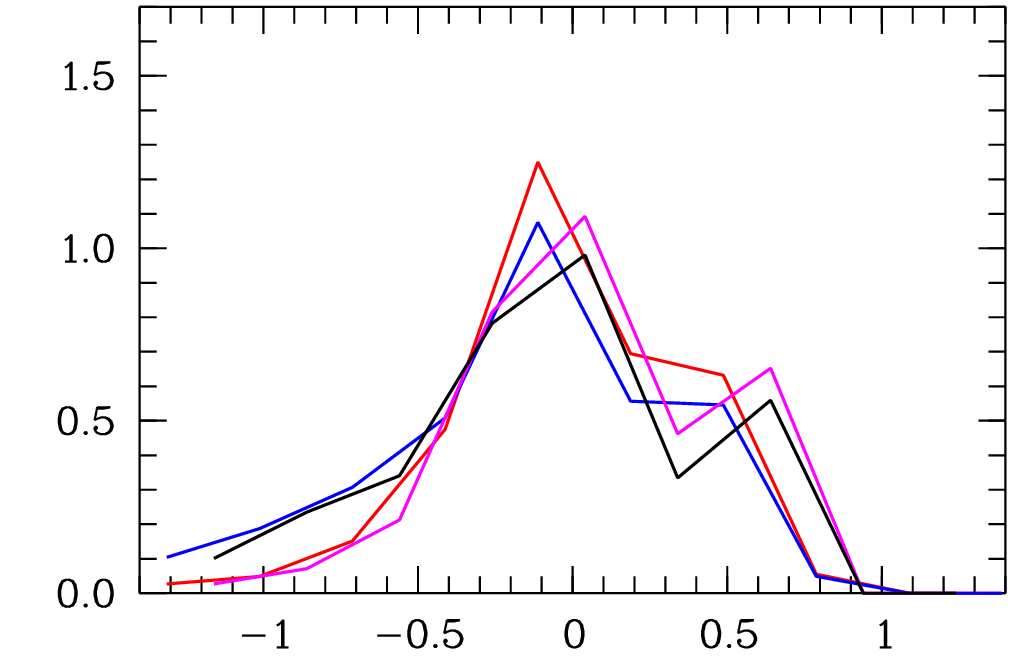}}\\
\resizebox{17cm}{!}{\includegraphics*{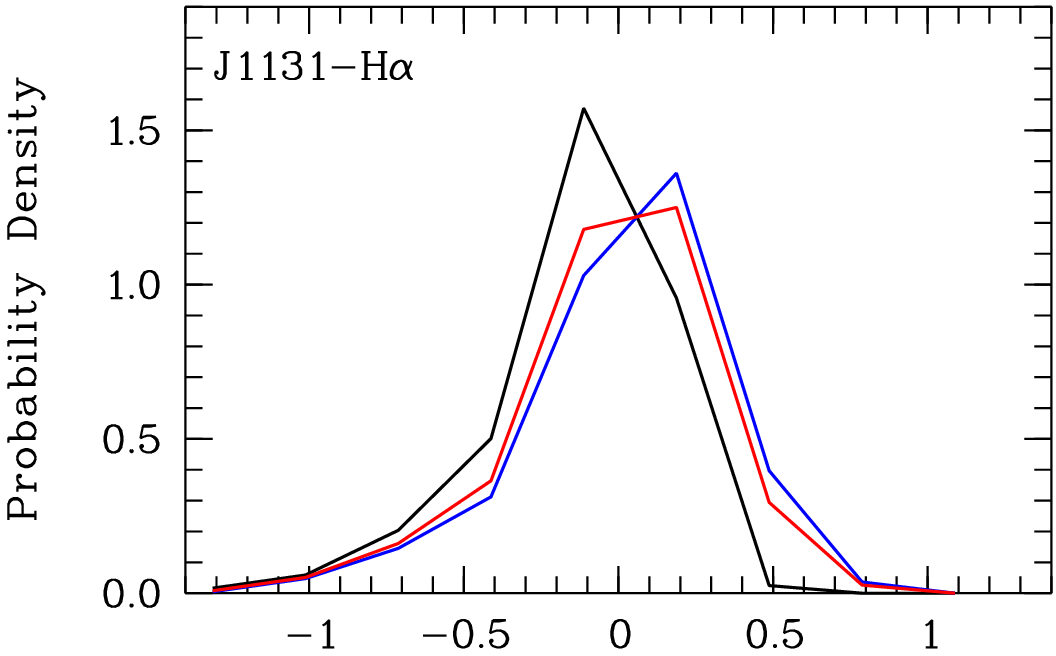}\includegraphics*{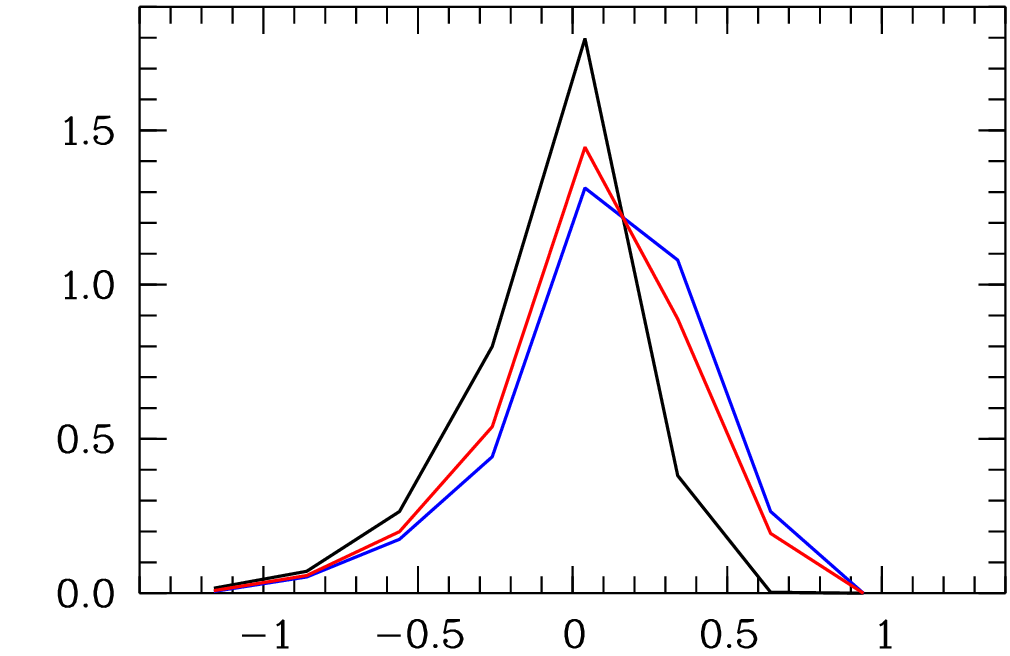}\includegraphics*{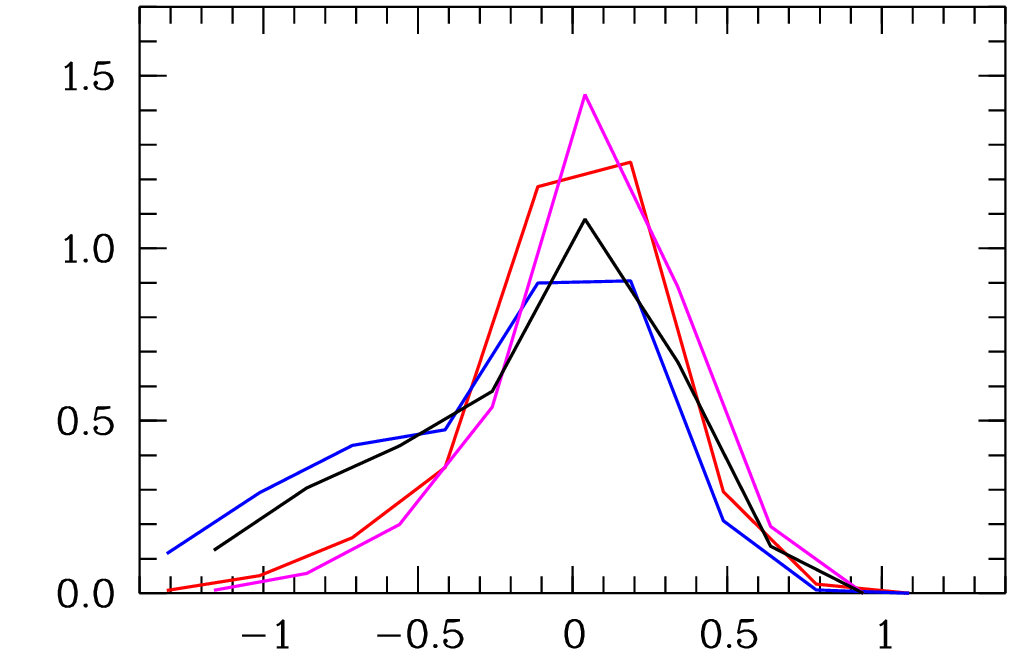}}\\
\resizebox{17cm}{!}{\includegraphics*{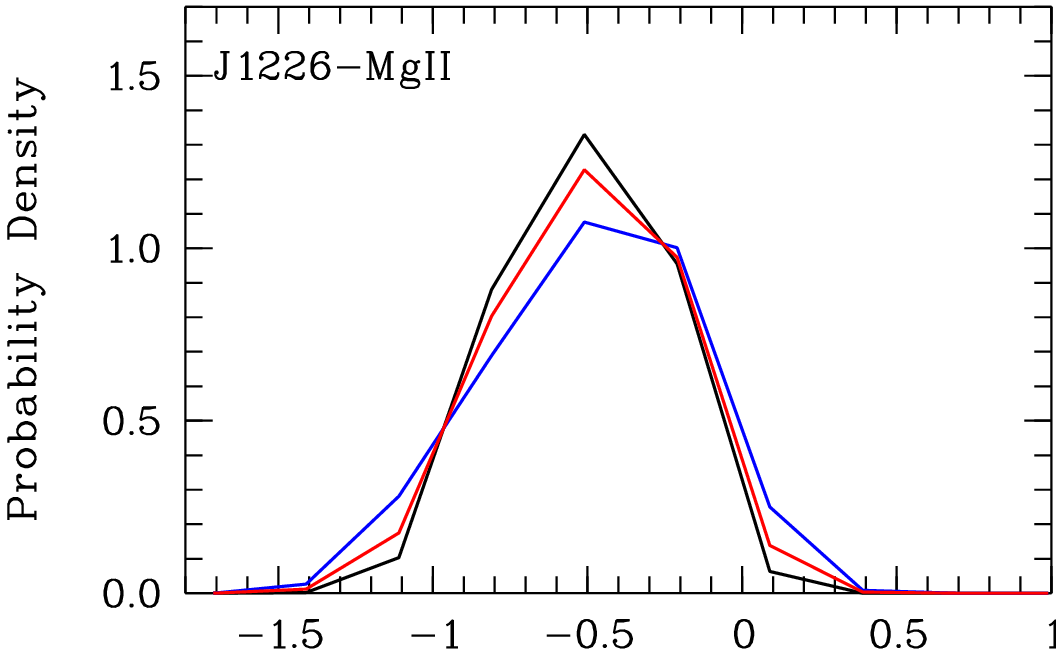}\includegraphics*{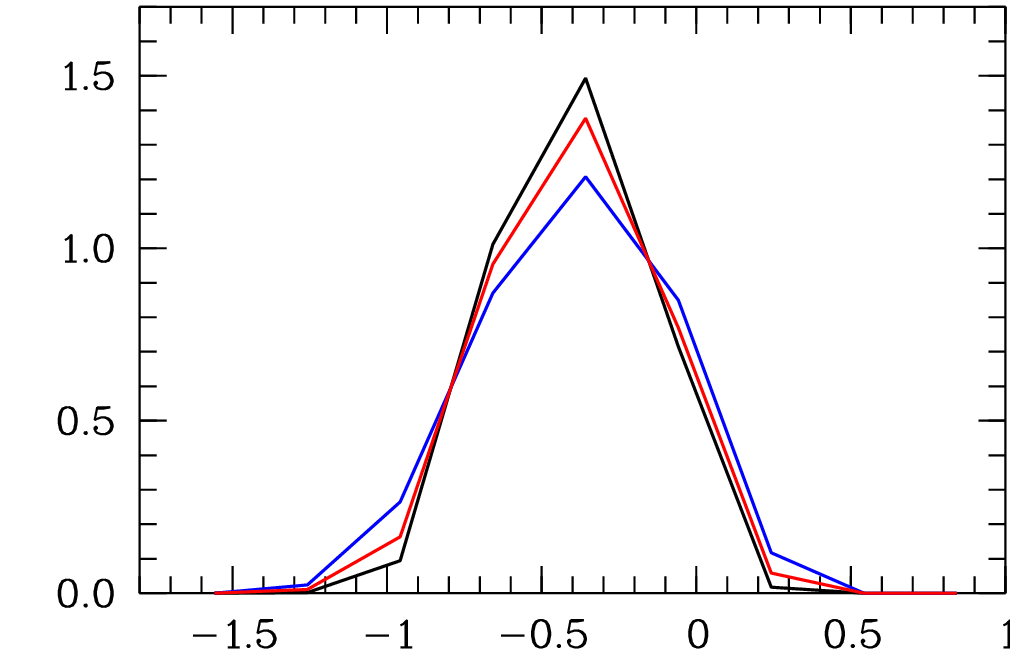}\includegraphics*{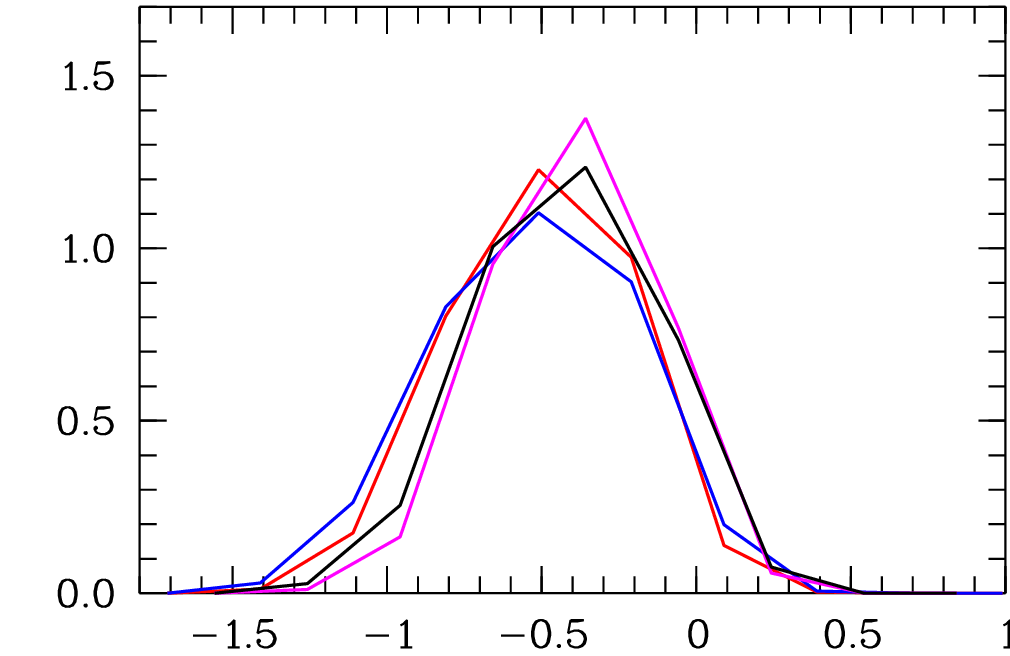}}\\
\resizebox{17cm}{!}{\includegraphics*{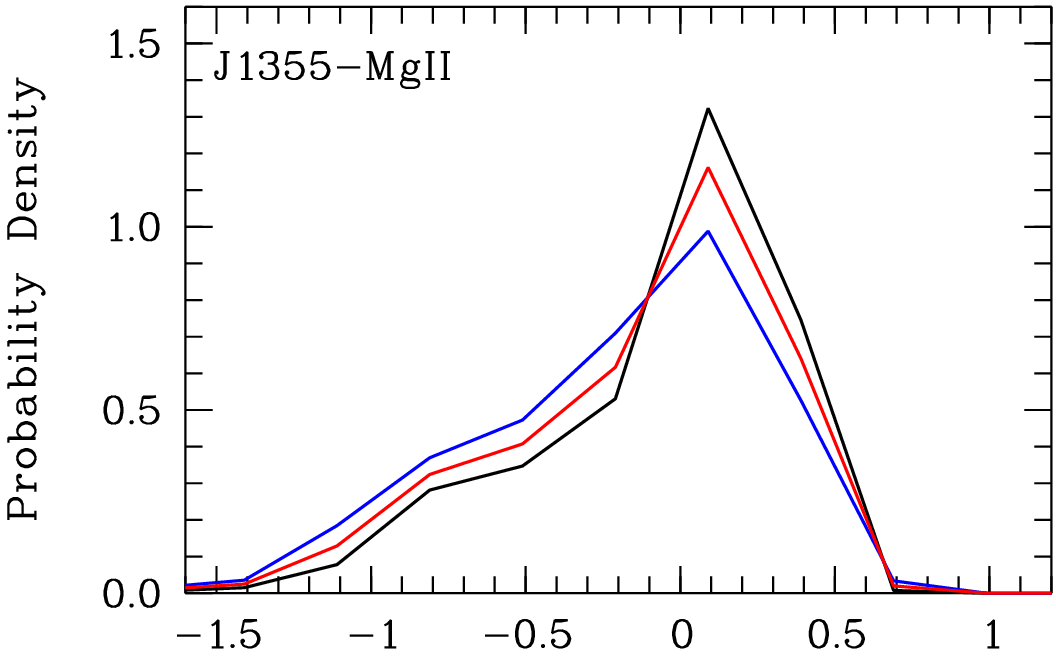}\includegraphics*{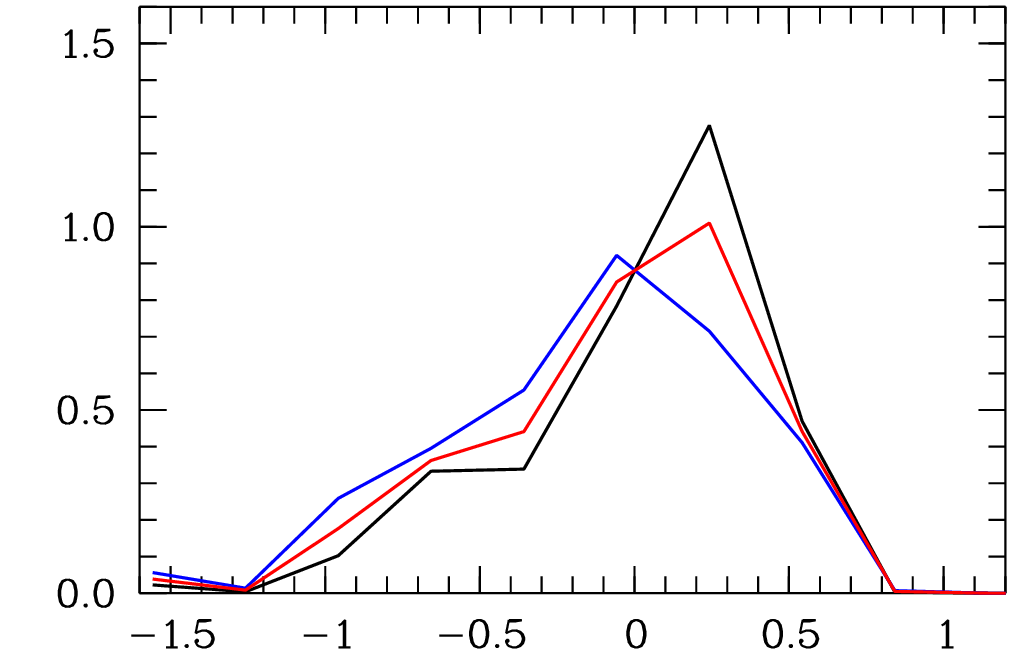}\includegraphics*{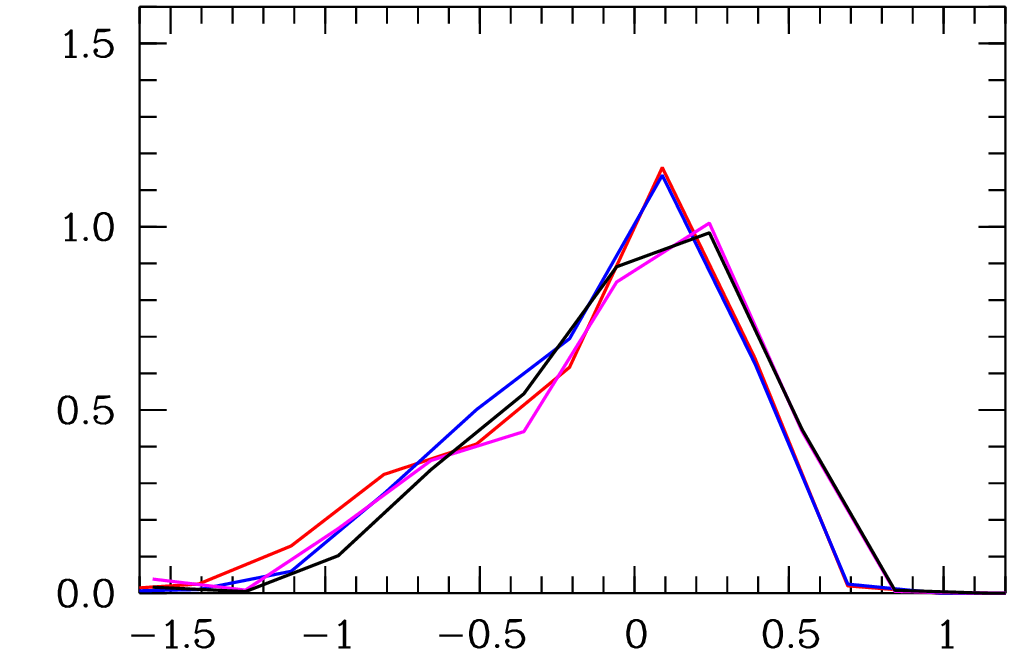}}\\
\resizebox{17cm}{!}{\includegraphics*{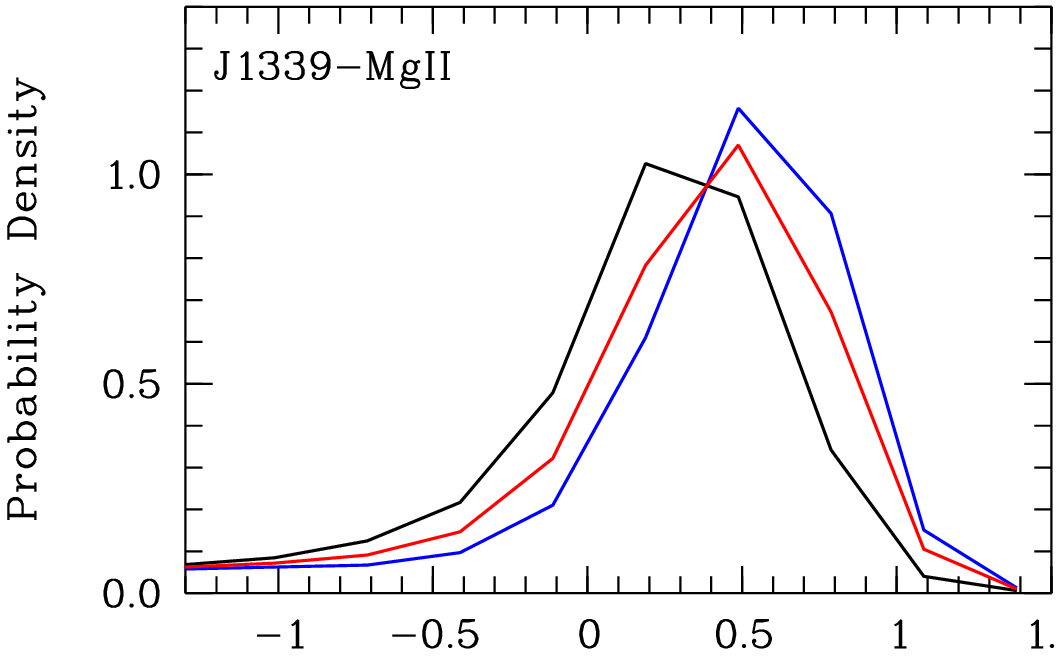}\includegraphics*{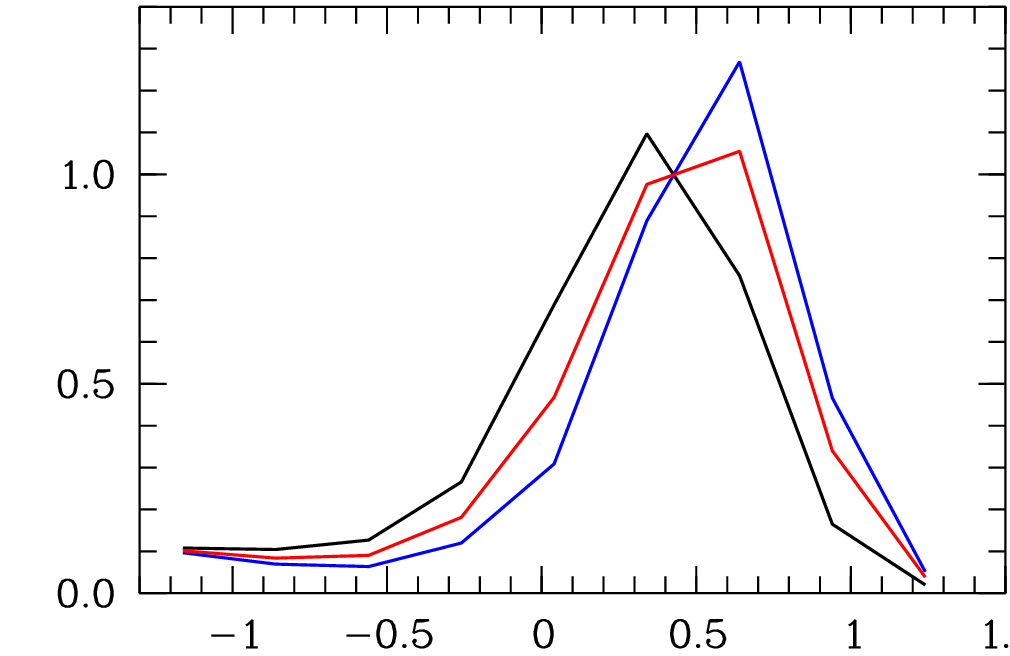}\includegraphics*{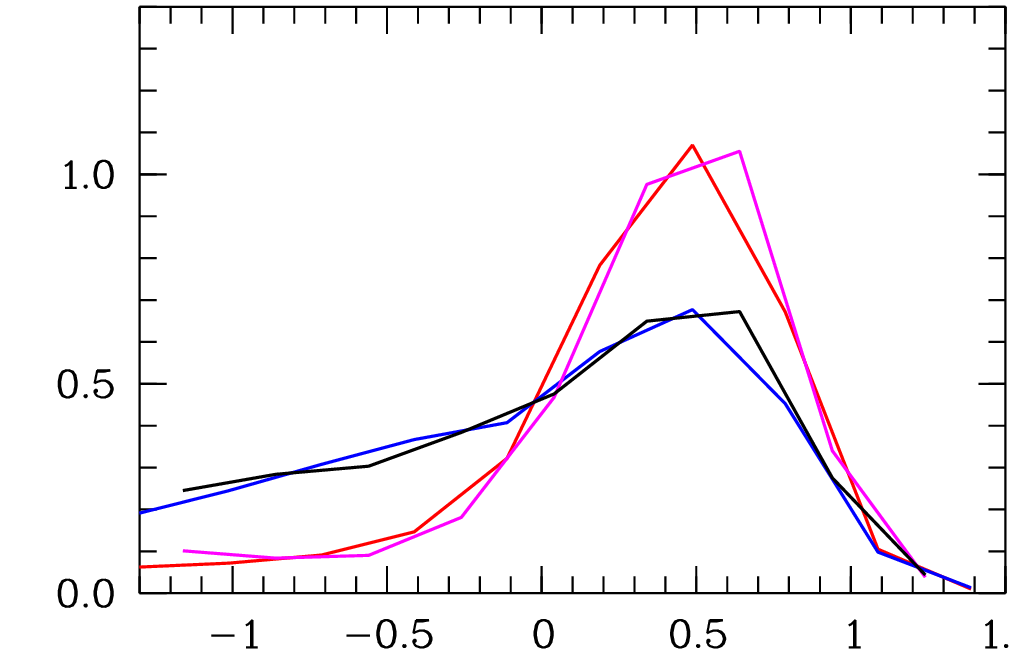}}\\
\resizebox{17cm}{!}{\includegraphics*{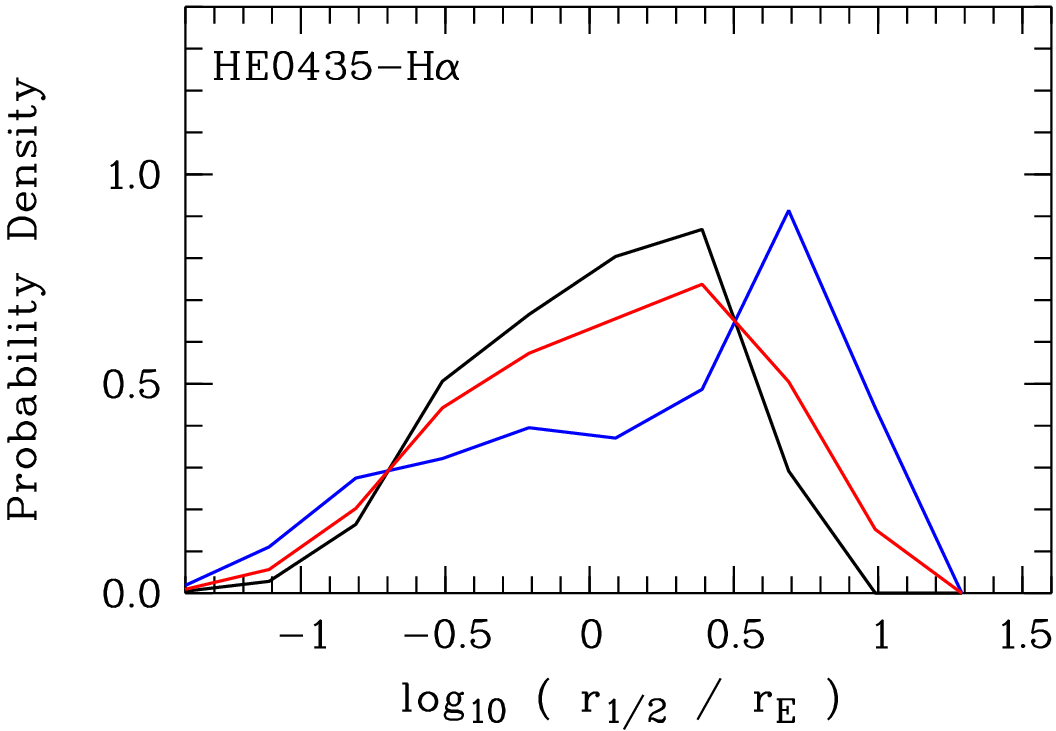}\includegraphics*{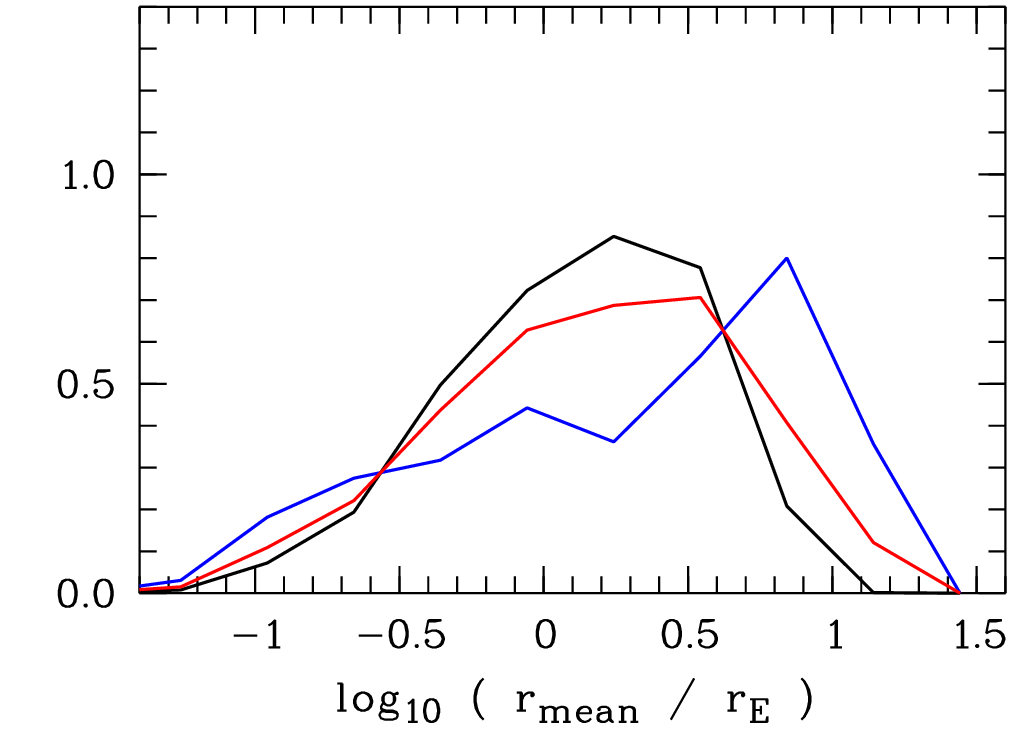}\includegraphics*{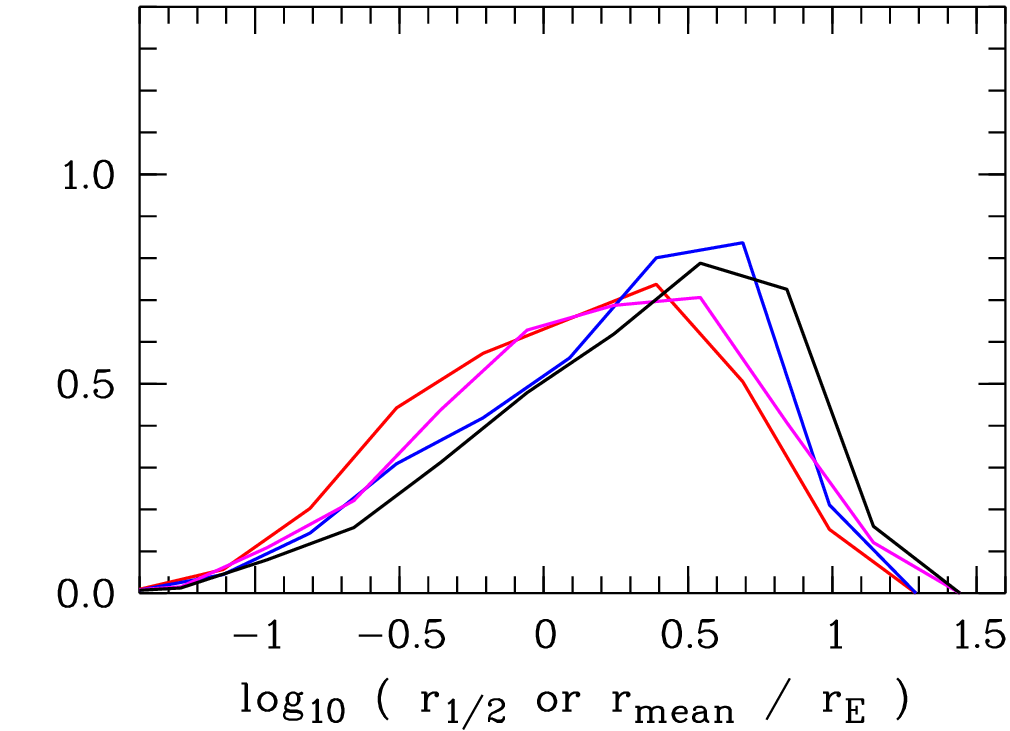}}
\caption{Posterior probability densities of the radius of the \ion{Mg}{ii} or H$\alpha$ BLR. The BLR radius is expressed in Einstein radius units. It was computed with $\mathcal{M} =  0.3  \mathcal{M}_{\odot}$, $r_{\text E}$ = 9.7, 9.2, 8.1, 10.1, 11.5 light-days for J1131, J1226, J1355, J1339, and HE0435, respectively. Left panels: Probability densities of the half-light radius, $r_{\text{1/2}}$, obtained with two magnification maps characterized by low (black) and high (blue) fractions of compact objects, and after marginalizing over the two maps (red). Middle panels: Same as the left panels but for the flux-weighted mean radius, $r_{\text{mean}}$. Right panels: Comparison of the probability densities, marginalized over the two maps, of the half-light radius (red and blue curves) and the flux-weighted mean radius (magenta and black curves), computed with constraints from the continuum source magnification (red and magenta curves) and without this constraint (blue and black curves).}
\label{fig:prblr}
\end{figure*}

\begin{table}[]
\caption{BLR radii}
\label{tab:radius}
\renewcommand{\arraystretch}{1.5}
\centering
\begin{tabular}{lcccc}
\hline\hline
 J1131 - \ion{Mg}{ii} &  $r_{\text{1/2}}$ & $r_{\text{mean}}$  \\
\hline
  Map $\kappa_{\star} / \kappa $ = 7\%   &     6.0    $^{+3.1}_{-2.5}$  &    6.1   $^{+3.8}_{-3.0}$  \\
  Map $\kappa_{\star} / \kappa $ = 30\%  &  12.2   $^{+13.9}_{-6.8}$  &   15.0  $^{+27.3}_{-9.0}$  \\
  All maps                              &      9.1   $^{+13.7}_{-4.7}$  &   10.2  $^{+24.3}_{-5.8}$  \\
  All maps; $\mu(v)$ fit only           &     7.5   $^{+13.2}_{-5.1}$  &    8.1  $^{+19.9}_{-5.4}$  \\
\hline
 J1131 - H$\alpha$ &  $r_{\text{1/2}}$ & $r_{\text{mean}}$  \\
\hline
  Map $\kappa_{\star} / \kappa $ = 7\%   &    8.4    $^{+4.3}_{-4.4}$  &    9.1   $^{+6.0}_{-4.7}$  \\
  Map $\kappa_{\star} / \kappa $ = 30\%  &   10.9    $^{+8.5}_{-5.7}$  &   12.2  $^{+10.5}_{-6.2}$  \\
  All maps                              &  9.9    $^{+7.7}_{-5.2}$  &   11.7   $^{+9.6}_{-6.1}$  \\
  All maps; $\mu(v)$ fit only           &   7.4    $^{+8.4}_{-5.6}$  &    8.6   $^{+9.4}_{-6.4}$  \\
\hline
 J1226 - \ion{Mg}{ii} &  $r_{\text{1/2}}$ & $r_{\text{mean}}$  \\
\hline
  Map $\kappa_{\star} / \kappa $ = 7\%   &    2.8    $^{+2.8}_{-1.1}$  &    3.6   $^{+2.9}_{-1.5}$  \\
  Map $\kappa_{\star} / \kappa $ = 21\%  &  3.0    $^{+3.3}_{-1.6}$  &    3.7   $^{+3.6}_{-1.9}$  \\
  All maps                              &    2.9    $^{+3.0}_{-1.3}$  &    3.7   $^{+3.2}_{-1.7}$  \\
  All maps; $\mu(v)$ fit only           &   2.8    $^{+3.3}_{-1.4}$  &    3.4   $^{+3.4}_{-1.7}$  \\
\hline
 J1355 - \ion{Mg}{ii} &  $r_{\text{1/2}}$ & $r_{\text{mean}}$  \\
\hline
  Map $\kappa_{\star} / \kappa $ = 7\%   &   9.1    $^{+7.4}_{-6.6}$  &   10.0   $^{+9.0}_{-7.3}$  \\
  Map $\kappa_{\star} / \kappa $ = 21\%  &   6.1    $^{+8.6}_{-4.6}$  &    6.6  $^{+10.3}_{-5.0}$  \\
  All maps                              &   7.7    $^{+8.1}_{-5.8}$  &    8.5   $^{+9.7}_{-6.5}$  \\
  All maps; $\mu(v)$ fit only           &   7.6    $^{+8.2}_{-5.2}$  &    8.3   $^{+9.8}_{-5.7}$  \\
\hline
 J1339 - \ion{Mg}{ii} &  $r_{\text{1/2}}$ & $r_{\text{mean}}$  \\
\hline
  Map $\kappa_{\star} / \kappa $ = 11\%  &   17.6   $^{+19.1}_{-11.6}$  &   19.7  $^{+23.4}_{-12.7}$  \\
  Map $\kappa_{\star} / \kappa $ = 52\%  &   29.8   $^{+29.7}_{-18.2}$  &   32.7  $^{+31.0}_{-20.0}$  \\
  All maps                              &  23.7   $^{+28.1}_{-15.3}$  &   26.6  $^{+30.2}_{-17.1}$  \\
  All maps; $\mu(v)$ fit only           &  13.3   $^{+30.8}_{-11.5}$  &   15.3  $^{+33.1}_{-13.0}$  \\
\hline
 HE0435 - H$\alpha$ &  $r_{\text{1/2}}$ & $r_{\text{mean}}$  \\
\hline
  Map $\kappa_{\star} / \kappa $ = 7\%   &  13.2   $^{+22.3}_{-9.1}$   &   17.0  $^{+24.7}_{-11.5}$     \\
  Map $\kappa_{\star} / \kappa $ = 21\%  &  25.7   $^{+52.0}_{-22.4}$   &   28.2  $^{+69.7}_{-24.3}$  \\
  All maps                              &  15.2   $^{+32.8}_{-11.3}$  &   18.2   $^{+36.8}_{-13.2}$    \\
  All maps; $\mu(v)$ fit only           &  24.0   $^{+39.2}_{-18.8}$  &  27.0  $^{+47.3}_{-20.2}$    \\
\hline
\end{tabular}
\tablefoot{The BLR radii are given in light-days, assuming an average microlens mass of 0.3  $\mathcal{M}_{\odot}$.}
\end{table}

\subsection{Size of the broad emission line region}
\label{sec:sizeblr}

We estimated the most likely BLR radius by marginalizing over all parameters but $r_{\text{in}}$. From the $r_{\text{in}}$ distribution, we computed the half-light radius,  $r_{\text{1/2}}$, and flux-weighted mean radius, $r_{\text{mean}}$, distributions, following \citet{2021Hutsemekers}. Figure~\ref{fig:prblr} shows the posterior probability densities, uniformly resampled on a logarithmic scale, of the BLR radii, $r_{\text{1/2}}$ and $r_{\text{mean}}$, for the different objects and lines. The results obtained with different values of the fraction of compact objects, $\kappa_{\star} / \kappa$, are illustrated separately. The probability distributions do not significantly depend on $\kappa_{\star} / \kappa$. The largest difference is observed for the \ion{Mg}{ii} BLR radius in J1131, but the distributions still largely overlap.

Our microlensing simulations simultaneously fit the magnification profile, $\mu(v)$, and the continuum magnification, $\mu^{cont}$ (Sect.~\ref{sec:models}). We also show, in the right panels of Fig.~\ref{fig:prblr}, the probability distributions computed by only taking into account the $\mu(v)$ profile (which necessitates much less computing time). In most cases, the probability distributions are in excellent agreement, indicating that taking into account $\mu^{cont}$ does not improve the constraints on the BLR size.  This can be explained by the fact that the BLR covers a larger area of the magnification map than the continuum source.

Table~\ref{tab:radius} gives the BLR radii, computed from the median values of the probability distributions shown in Fig.~\ref{fig:prblr}. The uncertainties correspond to the equal-tailed credible intervals that enclose a posterior probability of 68\%. The radii and intervals are then converted to a linear scale, multiplied by $r_{\text E}$, and expressed in light-days. To compute the Einstein radius (Eq.~\ref{eq:re}), we adopted a flat lambda cold dark matter ($\Lambda$CDM) cosmology with $H_0 = 68$ km~s$^{-1}$ Mpc$^{-1}$ and $\Omega_m$ = 0.31. We derived $r_{\text E}$ = 9.7, 9.2, 8.1, 10.1, 11.5 $\times \sqrt{ \mathcal{M} / 0.3  \mathcal{M}_{\odot}}$ light-days for J1131, J1226, J1355, J1339, and HE0435, respectively. $\mathcal{M}_{\odot}$ is the solar mass.

For a given quasar line, all of the BLR radii agree within the uncertainties. We found no significant difference between $r_{\text{1/2}}$ and $r_{\text{mean}}$, or between measurements with and without the constraint from $\mu^{cont}$. For J1131, we found no difference between the \ion{Mg}{ii} and H$\alpha$ BLR radii. On the other hand, the \ion{Mg}{ii} BLR radius in J1339 is about a factor of four larger than the radius of the \ion{C}{iv} BLR measured by \citet{2024Hutsemekers}. This is in agreement with the difference between the low- and high-ionization BLR radii reported by \citet{2013Guerras} and \citet{2021Fian}.

\subsection{Radius-luminosity relations}
\label{sec:radlum}

\begin{figure}[]
  \resizebox{0.95\hsize}{!}{\includegraphics*{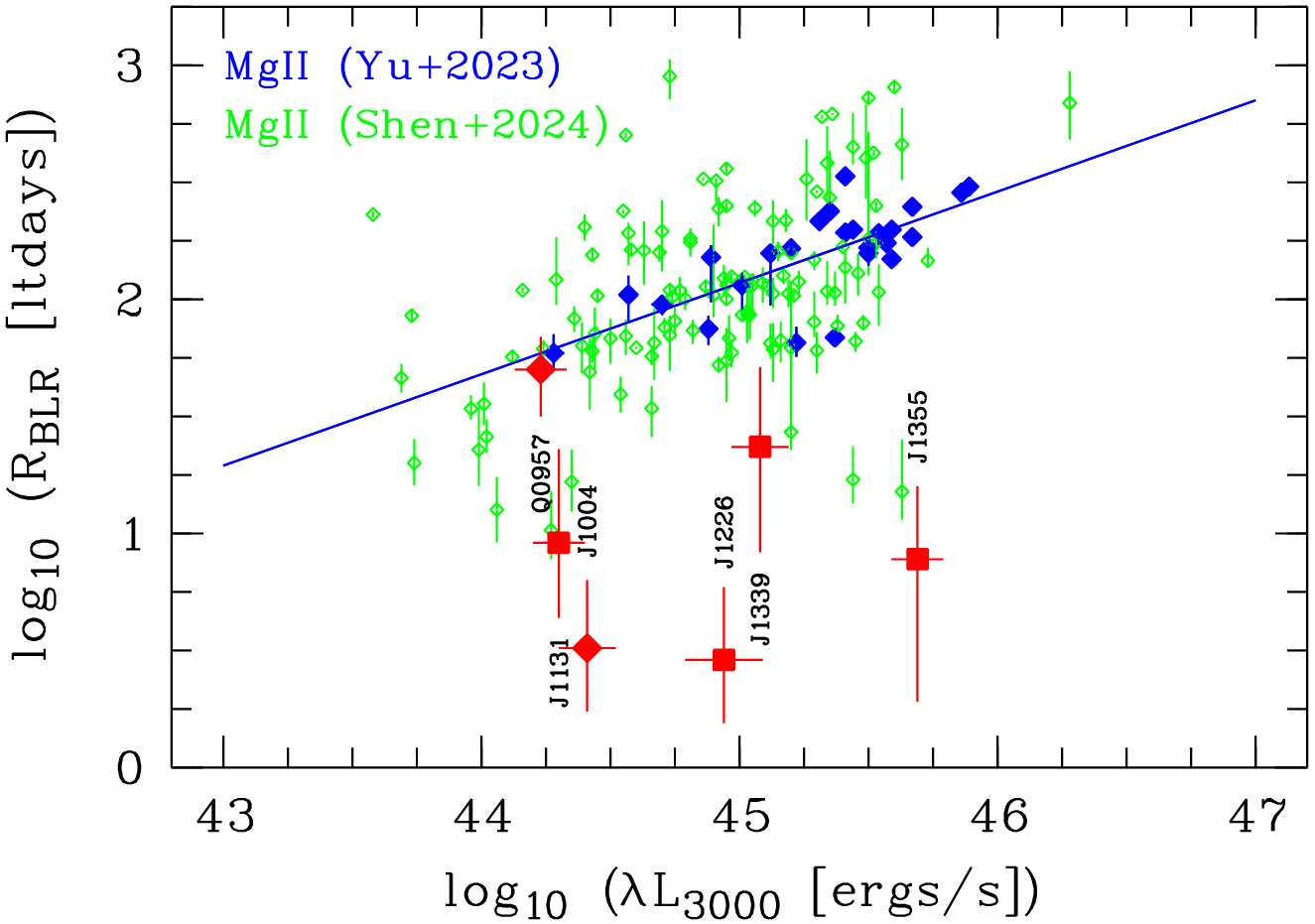}}
\caption{Radius--luminosity relation for the \ion{Mg}{ii} BLR. The rest-frame time lag from reverberation mapping and the continuum luminosity at 3000~\AA\ are from \citet[in blue]{2023Yu} and \citet[in green]{2024Shen}. The fit from \citet{2023Yu} is superimposed as a continuous line. The BLR half-light radii measured from microlensing are superimposed in red. Squares show the measurements from this work. Diamonds show the measurements from \cite{2023Fian,2024Fian} for Q0957$+$561 and J1004$+$4112.}
\label{fig:radlummgii}
\end{figure}

\begin{figure}[]
  \resizebox{0.95\hsize}{!}{\includegraphics*{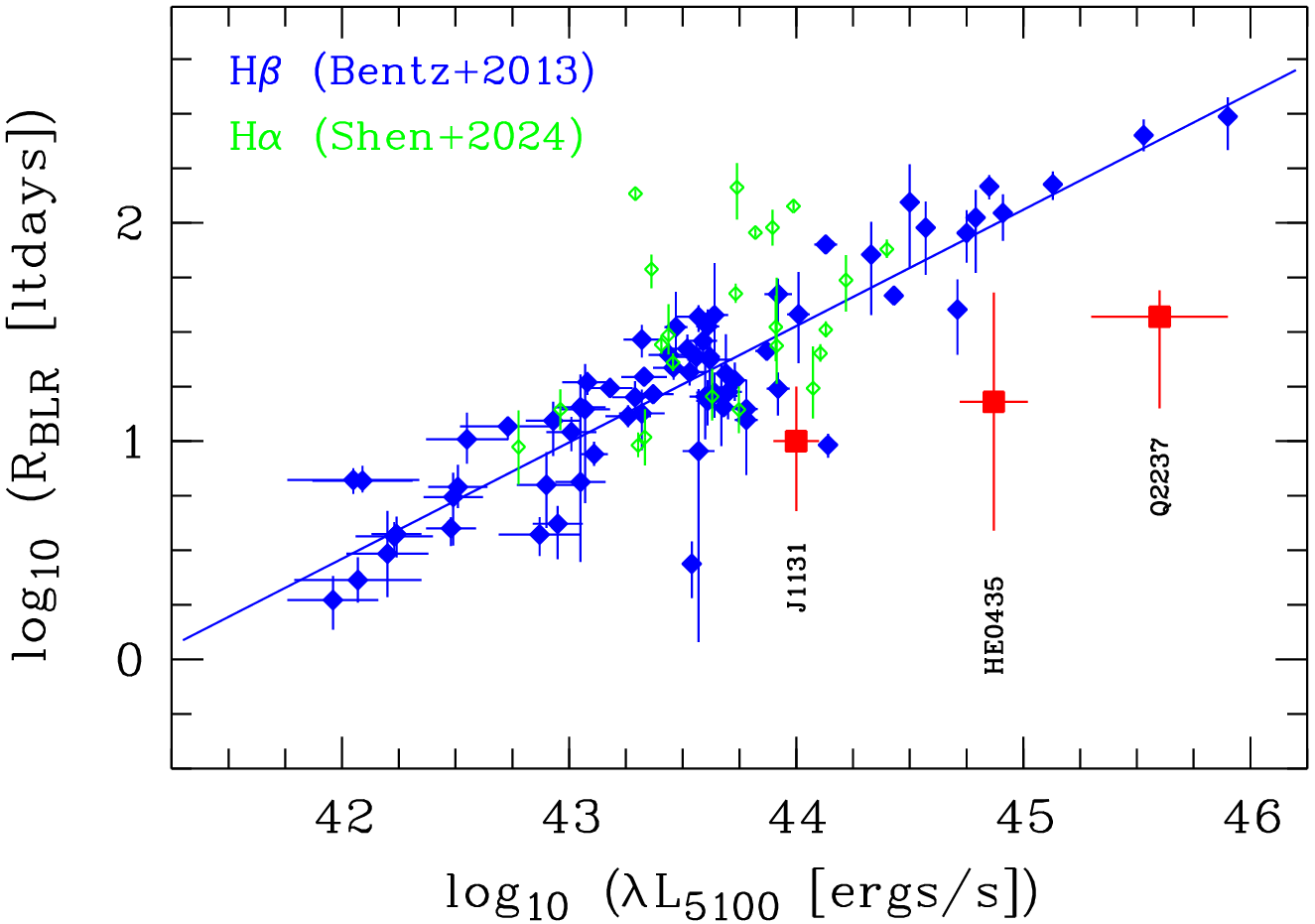}}
\caption{Radius--luminosity relation for the H$\alpha$ and H$\beta$ BLRs. The rest-frame time lag from reverberation mapping and the continuum luminosity at 5100~\AA\ are from \citet[H$\beta$ measurements, in blue]{2013Bentz} and \citet[H$\alpha$ measurements, in green]{2024Shen}. The fit from \citet{2013Bentz} is superimposed as a continuous line. The H$\alpha$ BLR half-light radii measured from microlensing are superimposed as red squares.}
\label{fig:radlumha}
\end{figure}

The $R-L$ relation for the \ion{Mg}{ii} BLR is illustrated in Fig.~\ref{fig:radlummgii}. It shows the reverberation mapping measurements of \citet{2023Yu} and \citet{2024Shen}, together with our measurements from microlensing (the ``All maps'' half-light radii reported in Table~\ref{tab:radius}). We added the \ion{Mg}{ii} BLR sizes independently obtained for Q0957$+$561 and J1004$+$4112 by \citet{2023Fian,2024Fian}. The $R-L$ relation for the H$\alpha$ BLR is illustrated in Fig.~\ref{fig:radlumha}. It includes the reverberation mapping measurements obtained for the H$\alpha$ BLR by \citet{2024Shen}. Since these data cover a limited luminosity range, we also showed the measurements of \citet{2013Bentz} for the H$\beta$ BLR, without any correction but keeping in mind that the H$\beta$ BLR could be smaller than the H$\alpha$ BLR by about 0.15 dex \citep{2017Grierb}. To the microlensing BLR radii obtained for J1131 and HE0435, we added the H$\alpha$ BLR radius of Q2237$+$0305 given in \citet{2024Savic}.
The macro-magnification-corrected luminosities are reported in Table~\ref{tab:Luminosities} for the five systems for which we derived a BLR size and for the three systems with measurements from literature; that is, Q0957$+$561,  J1004$+$4112, and Q2237$+$0305. When the monochromatic luminosities were not available for a wavelength of interest (i.e., 3000 \AA), we applied the bolometric correction ratios from \cite{2012Sluse} to convert published monochromatic luminosities to $\lambda L_{\lambda} (3000 \AA)$. For  Q0957$+$561, we used  $\lambda L_{\lambda} (5100 \AA)$ from \citet{2013Guerras} and assumed an uncertainty of 0.1~dex. For J1004$+$4112, we used $\lambda L_{\lambda} (1350 \AA)$ from  \cite{2023Hutsemekers}.

We immediately see from Figs.~\ref{fig:radlummgii} and~\ref{fig:radlumha} that, except for Q0957$+$561, the BLR radii derived from microlensing are systematically below the $R-L$ relations obtained with reverberation mapping. For the \ion{Mg}{ii} BLR, the deviation is particularly strong, reaching a factor of 30 for J1226.  A smaller deviation was already suspected for the \ion{C}{iv} $R-L$ relation \citep{2024Hutsemekers}.

As was shown in the previous section, the microlensing BLR radii are robust, within uncertainties, with respect to changes in the magnification map, in particular the stellar mass fraction, and to changes in the radius calculation, either half-light or flux-weighted. As is discussed in \citet{2024Hutsemekers}, the microlensing BLR radius is sensitive to the macro-magnification factor used in the computation of $\mu(v)$. For J1226, we then considered a higher, but still reasonable value: $M$ = 1.4$\pm$0.1 such that $\mu(v \! = \! 0) \simeq 1$. With this value, we obtained $r_{\text{1/2}}$ = 4.6 $^{+4.2}_{-2.6}$ light-days instead of 2.9$^{+3.0}_{-1.3}$ light-days, which is slightly higher and within the uncertainties, but still far from a factor of 30. We also computed $\mu(v)$ as the ratio of the magnification profiles produced with maps specific to images A and B, and thus have assumed a significant microlensing effect in image B. We found $r_{\text{1/2}}$ = 3.3 $^{+7.0}_{-1.7}$  light-days, which is again in good agreement with the values reported in Table~\ref{tab:radius}. The same exercise with J1355 leads to $r_{\text{1/2}}$ = 9.0 $^{+13.7}_{-6.6}$ light-days, which is equal, within the uncertainties, to the value given in Table~\ref{tab:radius}. Finally, a change in the average stellar mass, $\mathcal{M} =  0.3 \mathcal{M}_{\odot}$, could in principle increase  $r_{\text{1/2}}$ but it would require an average stellar mass that is almost three orders of magnitude larger to explain a factor of 30, which is unrealistic \citep{2010Poindextera,2019Jimenez}.

\begin{table}[]
\caption{Magnification-corrected monochromatic luminosities. }
\label{tab:Luminosities}
\renewcommand{\arraystretch}{1.2}
\centering
\begin{tabular}{lccc}
\hline\hline
                Object &   $\log[\lambda L_{\lambda} (3000 \AA)]$  &  $\log[\lambda L_{\lambda} (5100 \AA)]$ & Ref.  \\
\hline
                HE0435 &   - & 44.87$\pm$0.15 & (1)  \\
                J1131        & 44.30$\pm$0.10     & 44.02$\pm$0.10 & (1)  \\  
                J1226        & 44.94$\pm$0.15 & - & (1)  \\
                J1355        & 45.70$\pm$0.10    & - & (1)   \\
                J1339$^{\dagger}$    &   45.08$\pm$0.11 & - & (2)  \\           
                Q2237$^{\dagger\dagger}$ & - & 45.6$\pm$0.3 & (1, 3) \\ 
                J1004 & 44.41$\pm$0.11 &  - & (4) \\ 
                Q0957  & 44.23$\pm$0.10 & - & (5) \\
\hline
\end{tabular}
\tablefoot{(1)  \cite{2012Sluse},  (2) \cite{2021Shalyapin},  (3)  \cite{2011Assef}, (4)  \cite{2023Hutsemekers},  (5) \cite{2013Guerras}. $\dagger$:~using a macro-magnification for image A of $9.4 \pm 2.2$ (see Sects. 3 and 4 of (2)) ; $\dagger\dagger$: average of values from (1) and (3). The luminosities are given in erg~s$^{-1}$.}
\end{table}

Based on a dense spectroscopic monitoring secured between 2008 and 2011 (about 150 spectra), we were able to estimate the size of the BLR in J1131 with the reverberation mapping technique (Sluse et al., in prep.). Using two different methods for measuring the time lag between the $R-$band continuum and the H$\beta$  emission, a BLR radius of about 15 light-days was obtained, with an uncertainty of around 30\%. Hence, the reverberation mapping radius is in agreement, within the uncertainties, with the BLR size estimated from single-epoch microlensing (Table~\ref{tab:radius}). Both techniques, microlensing and reverberation mapping,\footnote{It should be kept in mind that, in some cases, the response-weighted time delay measured with reverberation mapping could differ from the flux-weighted radius by up to a factor of two \citep[][]{2024Rosborough}.} thus give a BLR size for J1131 below the $R-L$ relation. While the measurement of the BLR size using both microlensing and reverberation mapping is only available for one object, this result further supports the reliability of the microlensing method in estimating BLR sizes.

Our measurements thus confirm that the intrinsic dispersion of the BLR radii with respect to the $R-L$ relation is large, as has already been shown by \citet[][]{2024Shen}. However, the fact that the microlensing BLR radii are systematically below the $R-L$ relations remains to be explained.

\begin{figure}[]
  \resizebox{0.95\hsize}{!}{\includegraphics*{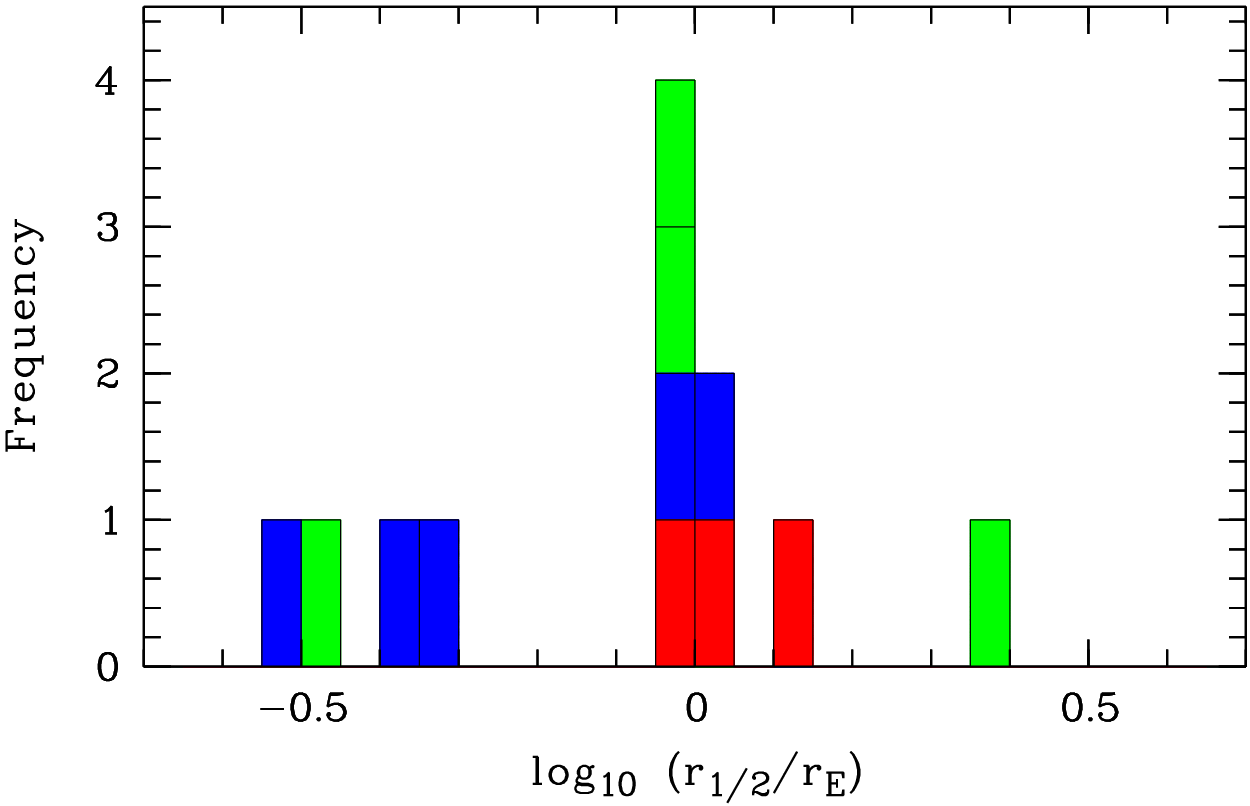}}
\caption{Distribution of the BLR half-light radii derived from our microlensing analysis, expressed in units of $r_{\text E}$. The \ion{Mg}{ii} and H$\alpha$ BLR radii from this work are represented in green and red, respectively, and the \ion{C}{iv} BLR radii from \citet{2023Hutsemekers,2024Hutsemekers} and \citet{2024Savic} in blue.}
\label{fig:histrblr}
\end{figure}
\begin{figure}[]
  \resizebox{0.95\hsize}{!}{\includegraphics*{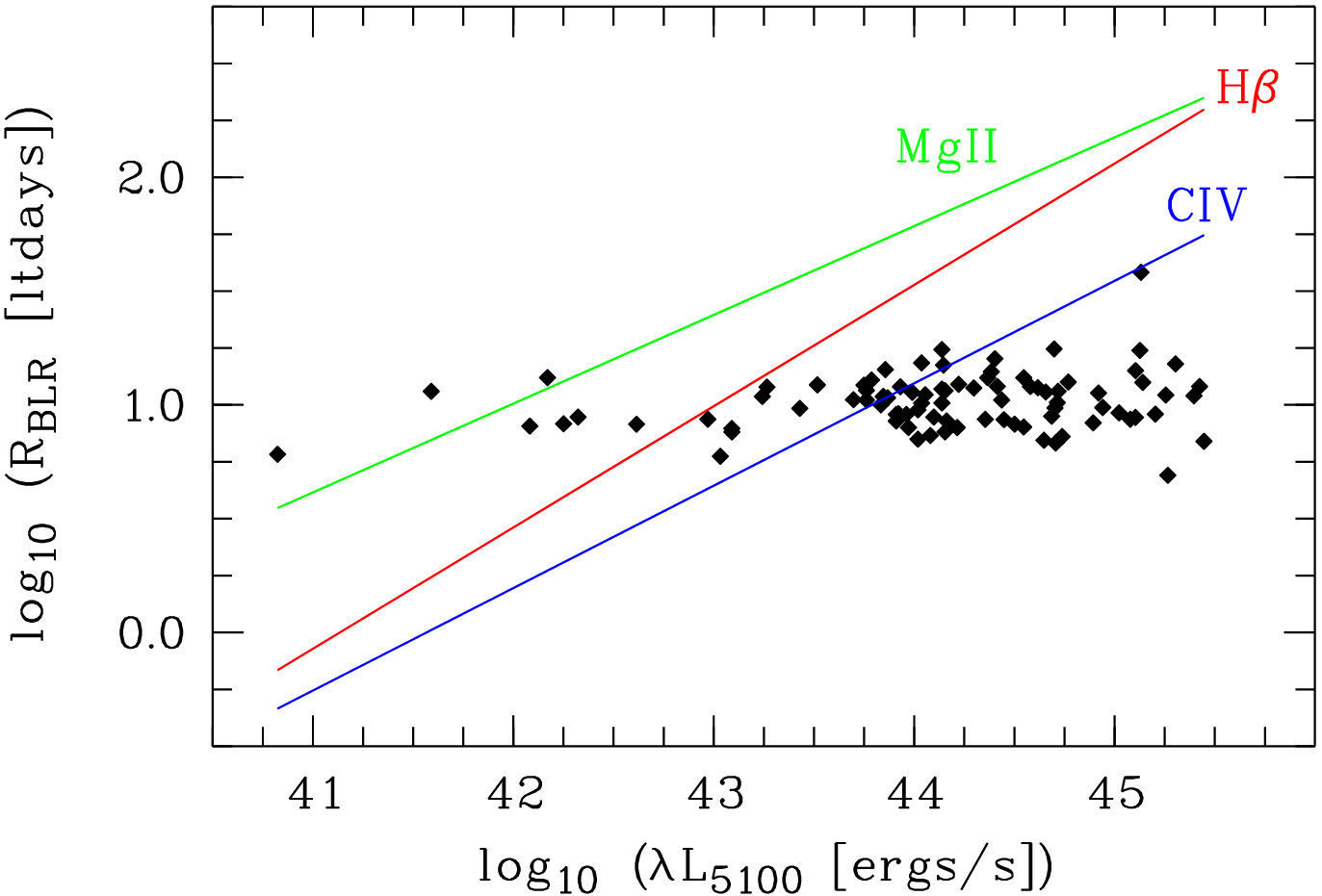}}
\caption{BLR radius -- quasar luminosity relations shown as a function of the luminosity at 5100~\AA\ (continuous lines) for the \ion{C}{iv} \citep[][in blue]{2021Kaspi}, \ion{Mg}{ii} \citep[][in green]{2023Yu}, and H$\beta$ \citep[][in red]{2013Bentz} BLRs. The microlensing Einstein radii of lensed quasars from the compilation of \citet{2011Mosquerab} are superimposed (black diamonds).}
\label{fig:radlummosquera}
\end{figure}

In Fig.~\ref{fig:histrblr}, we show the distribution of the microlensing BLR radii expressed in Einstein radii, for the quasars studied so far. The BLR radius is comparable to the Einstein radius within roughly a factor of two (0.3 dex). This can be expected, since a very small BLR with respect to the Einstein radius would be entirely magnified, with no line profile distortion, while a much larger BLR would be totally unaffected. This distribution is the result of our selection of lensed quasars that show line profile distortions. In Fig.~\ref{fig:radlummosquera}, assuming $R_{\text{BLR}} \simeq r_{\text E}$, we plotted the microlensing Einstein radius of lensed quasars against the quasar luminosity using the sample compiled by \citet{2011Mosquerab}. The luminosities at 5100~\AA\  were computed back using the BLR radius and the $R-L$ relation given in that paper. With these luminosities and the bolometric corrections from \citet{2012Sluse}, we derived the luminosities at 1350~\AA\ and 3000~\AA, from which we computed the $R-L$ relations for H$\beta$ \citep{2013Bentz}, \ion{Mg}{ii}  \citep{2023Yu}, and \ion{C}{iv} \citep{2021Kaspi}, as a function of the 5100~\AA\ luminosity. These relations are shown in  Fig.~\ref{fig:radlummosquera}.

If we pick objects with observed line profile distortions, and hence $R_{\text{BLR}} \simeq r_{\text E}$, we see that the BLR radii of high-luminosity ($\log[\lambda L_{\lambda} (5100 \AA)] \gtrsim 44$) quasars are below the $R-L$ relations. The selection effect appears stronger for \ion{Mg}{ii}, intermediate for H$\alpha$, and smaller for \ion{C}{iv}, in agreement with the deviations observed in Figs.~\ref{fig:radlummgii} and~\ref{fig:radlumha}, and in Fig.~8 of \citet{2024Hutsemekers}, where the average deviations are measured to be 1.0, 0.64, and 0.48 dex, for \ion{Mg}{ii}, H$\alpha$, and \ion{C}{iv}, respectively. This selection bias appears because the microlensing Einstein radii of currently detected lensed quasars are strongly concentrated around a single value, about ten light-days for an average microlens mass of 0.3  $\mathcal{M}_{\odot}$. As can be calculated from  Eq.~\ref{eq:re}, this follows from the fact that the lens redshift distribution peaks around $z_l \simeq 0.5 - 1.0$ for lensed quasars at $z_s \simeq 2$ \citep[e.g.,][]{2003Ofek}.

The fact that the microlensing BLR radii are systematically below the $R-L$ relations is thus a selection effect.
Since the selection bias depends on the line under study, it should also be taken into account when comparing, on a statistical basis, the sizes of the regions that emit different lines. In particular, the difference between the averaged \ion{Mg}{ii} and \ion{C}{iv} BLR sizes would be underestimated if selection bias corrections were not applied.

\section{Summary and conclusions}
\label{sec:conclusions}

We have analyzed the microlensing-induced line profile distortions observed in \ion{Mg}{ii} and/or H$\alpha$ in five gravitationally lensed quasars, J1131-1231, J1226-0006, J1355-2257, J1339+1310, and HE0435-1223, using single-epoch spectroscopic data. This study complements our previous analyses of the \ion{C}{iv} high-ionization line distortions observed in four lensed quasars, Q2237+0305, J1004+4112, J1339+1310, and J1138+0314 \citep{2021Hutsemekers,2023Hutsemekers,2024Hutsemekers,2024Savic}.
We first noticed that there are a wide variety of $\mu(v)$ magnification profiles. While the line core is often less magnified, the blue and red wings can either both be magnified, both be demagnified, or one can be magnified and the other one demagnified, or they can simply be unaffected. For a given object, the $\mu(v)$ profiles of different lines have roughly similar shapes, although the high-ionization line profiles tend to show stronger magnifications than the low-ionization ones.

To characterize the size, geometry, and kinematics of the BLR, we have compared the observed line profile distortions with simulated ones. The simulations are based on three BLR models (KD, EW, and PW) of different sizes, inclinations, and emissivities, convolved with microlensing magnification maps specific to the microlensed quasar images. We found that:
\begin{itemize}
\item The observed line profile distortions can all be reproduced with microlensing-induced distortions of line profiles generated from our simple BLR models. In several cases, the observed distortions can be simulated using either the KD or EW models, depending on the orientation of the magnification map with respect to the BLR axis. This demonstrates that the $\mu(v)$  magnification profiles depend on the position and orientation of the isovelocity parts of a BLR model with respect to the caustic network, and not only on their different effective sizes.
\item For J1131, J1226, and HE0435, the most likely model for the \ion{Mg}{ii} and H$\alpha$ BLRs is either KD or EW, depending on the map orientation relative to the BLR axis.  For the \ion{Mg}{ii} BLR in J1355 and J1339, the EW model is preferred. For all objects, the PW model has a lower probability. The inclination is poorly constrained, but it is smaller than 45\degr\ , as is expected for type~1 quasars. As for the high-ionization \ion{C}{iv} BLR, we can conclude that disk geometries with kinematics dominated by either Keplerian rotation or equatorial outflow best reproduce the microlensing effects on the low-ionization \ion{Mg}{ii} and H$\alpha$ emission line profiles.
\item The half-light radii of the \ion{Mg}{ii} and H$\alpha$ BLRs were measured in the range of 3 to 25 light-days. In J1131, the \ion{Mg}{ii} and H$\alpha$ BLRs have the same radius. In J1339, the \ion{Mg}{ii} BLR radius is about a factor of four larger than the radius of the \ion{C}{iv} BLR. The measured BLR radii are robust, within uncertainties, to changes in the magnification map, in particular the stellar mass fraction, and to changes in the radius calculation. For J1131, the BLR radius derived from reverberation mapping is in agreement with the microlensing BLR radius within the uncertainties.
\item The microlensing BLR radii of the \ion{Mg}{ii} and H$\alpha$ BLRs were compared with $R -L$ relations derived from reverberation mapping, and found to be systematically below. For the \ion{Mg}{ii} BLR, the deviation is particularly strong, reaching a factor of 30 for J1226. This confirms that the intrinsic dispersion of the BLR radii with respect to the $R-L$ relations is large. This also reveals a strong selection bias that affects microlensing-based BLR size measurements. This bias arises from the fact that, if microlensing-induced line profile distortions are observed in a lensed quasar, the BLR radius should be comparable to the Einstein radius, which clusters in a narrow range of values regardless of the (realistic) redshifts of the lens and the source.
\end{itemize}

\begin{acknowledgements}
D.H. and Đ.S. acknowledge support from the Fonds de la Recherche Scientifique - FNRS (Belgium) under grants PDR~T.0116.21 and No 4.4503.19.
\end{acknowledgements}

\bibliographystyle{aa}
\bibliography{references}

\begin{appendix}

\section{Size is not everything in BLR microlensing}
\label{sec:notonlysize}

Recently \citet{2024Fianc} used the \ion{Si}{iv} and \ion{C}{iv} magnification profiles in five lensed quasars to infer the inner BLR kinematics. They essentially showed that the magnification profiles can be reproduced with the KD kinematics, in agreement with results reported in \citet{2021Hutsemekers}, \citet{2023Hutsemekers,2024Hutsemekers}, and \citet{2024Savic}. However, they rejected alternatives that involve radial motions, arguing that precise fine-tuning would be needed to reproduce the observed magnification profiles. On the contrary, we have shown that both the KD and EW models can reproduce the same magnification profiles with similar probabilities, depending on the orientation of the BLR axis with respect to the caustic map. This dependence on the BLR axis - magnification map relative orientation is due to the fact that in KD models the highest velocities come from regions located, in projection, perpendicular to the BLR axis, while in EW models these regions are parallel to the BLR axis \citep[see Fig.~3 of][]{2017Braibant}. Furthermore, in the specific case of J1339$+$1310, we showed that the EW model reproduces the \ion{C}{iv} magnification profile better than the KD models. Similar conclusions are reached for the low-ionization lines studied in the present paper.

\begin{figure}[t]
  \resizebox{0.85\hsize}{!}{\includegraphics*{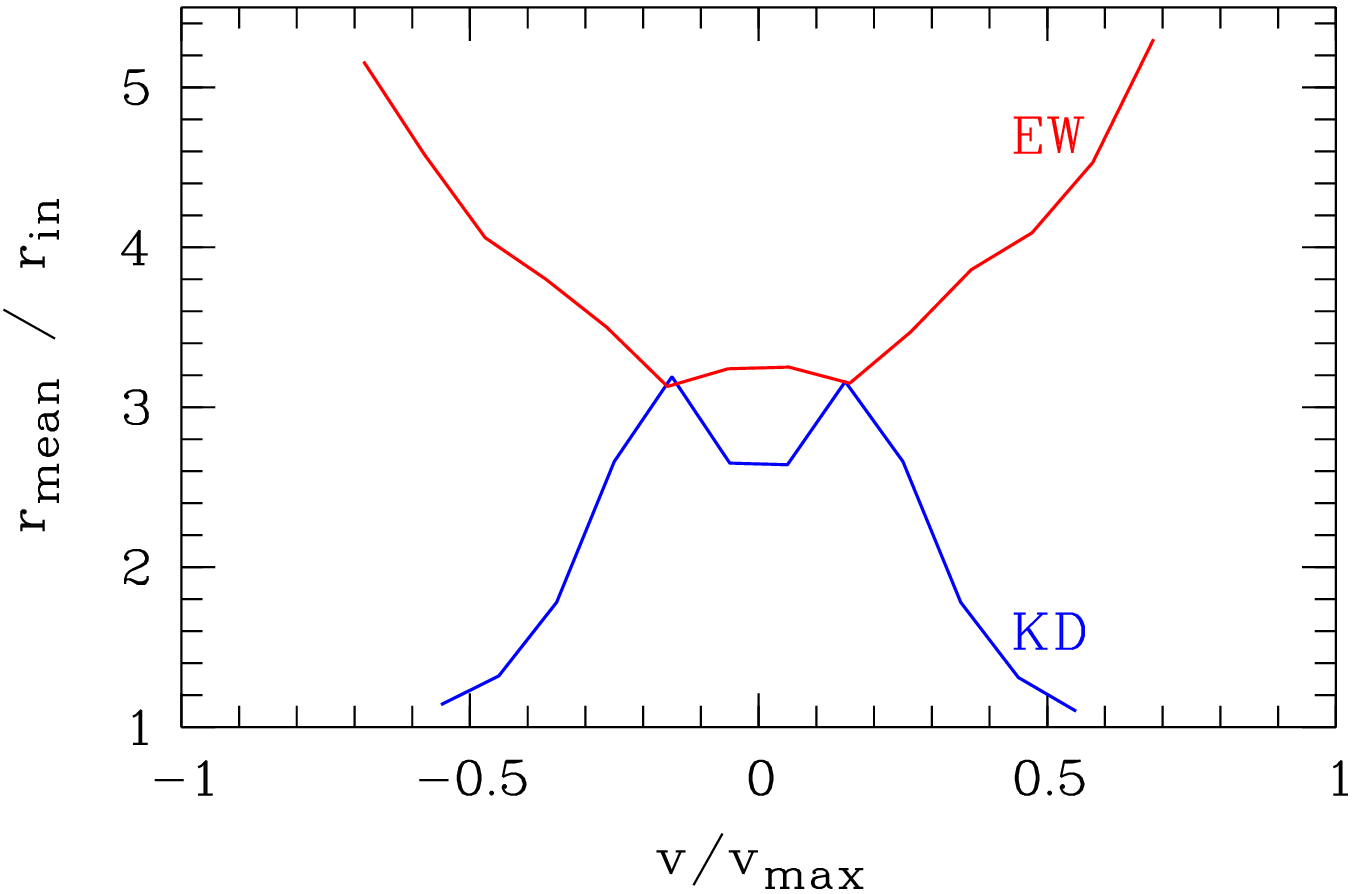}}%
  \caption{Flux-weighted mean radius of the BLR isovelocity slices as a function of the Doppler velocity in the line profile. The mean radius is computed for the 20 isovelocity slices of the KD and EW models, with an emissivity $\varepsilon = \varepsilon_0 \, (r_{\text{in}}/r)^{3}$, and seen at an inclination of 34\degr. A more complete version can be found in \citet{2017Braibant}. Using the half-radius instead of the mean radius does not change the velocity dependence.}
\label{fig:radii}
\end{figure}

\begin{figure}[t]
  \resizebox{0.9\hsize}{!}{\includegraphics*{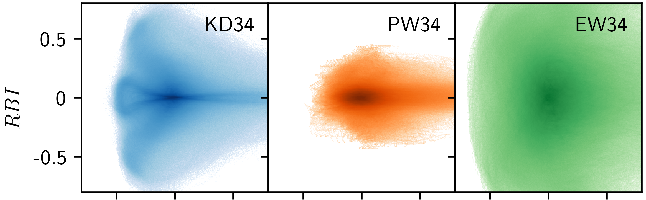}}\\%
  \resizebox{0.9\hsize}{!}{\includegraphics*{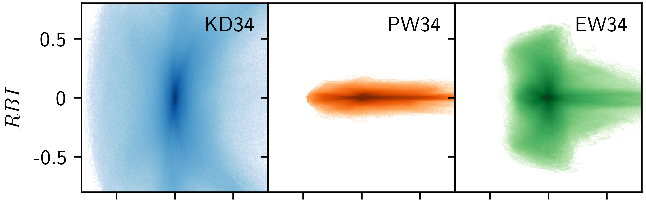}}\\%
  \resizebox{0.9\hsize}{!}{\includegraphics*{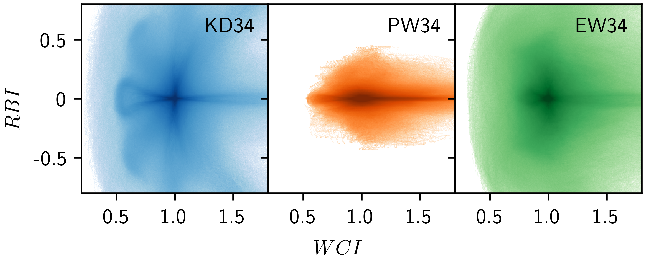}}%
  \caption{Two-dimensional histograms of simulated WCI and RBI (the darkest regions correspond to the highest frequencies).  These indices were measured from simulated line profiles that arise from the BLR models KD, PW, and EW seen at an inclination of 34\degr, and with an emissivity $\varepsilon = \varepsilon_0 \, (r_{\text{in}}/r)^{3}$. The simulations include the map orientations $\theta$ = [0\degr, 30\degr] (top panel), $\theta$ = [60\degr, 90\degr] (middle panel), and  $\theta$ = [0\degr, 30\degr, 60\degr, 90\degr] (bottom panel) which is the combination of the top and middle panels. These histograms were built from the magnification maps used for the analysis of J1004$+$4112 \citep{2023Hutsemekers}. They show a particularly clear dependence on the map orientation. The simulations were restricted to $0.9 \leq \mu^{BLR} \leq 1.1$, and to a single value of the BLR size, $r_{\text{in}} = 0.1 \, r_{\text E}$.}
\label{fig:hist2}
\end{figure}

The analysis of \citet{2024Fianc} is based on the hypothesis that the magnification of a velocity bin in the line profile is only determined by the size of the BLR subregion at the origin of this velocity bin. In this framework, since the observed (de)magnification is often stronger at large velocities, the size of the emitting region is assumed to decrease with increasing velocity, which can be explained by the KD kinematics. In our analysis, we do not make this assumption; the relative positions and sizes of the BLR subregions are fixed within the adopted model, and cannot be arbitrarily assigned. The emissivity law, in particular, contributes to constrain the size of the emitting regions.  The location and orientation of the different BLR subregions with respect to the caustic network are then found as important as the size to reproduce the observed magnification profiles, a conclusion already reached by \citet{2007Abajas} for a biconical geometry. Without the hypothesis that only size controls the magnification, both the KD and EW models can fit the same observed magnification profiles (see Fig.~\ref{fig:fitmuv}, and Fig.~7 of \citealt{2023Hutsemekers}), even if the mean emission radius increases as a function of the velocity in the EW model (Fig.~\ref{fig:radii}). Note also that the mean emission radius of the KD models does not change by more than a factor of three as a function of the velocity (and by no more than about 70\% for the EW model), so that size alone cannot explain the high differential magnifications observed in the BELs of some objects. In Fig.~\ref{fig:hist2} \citep[see also][for other examples]{2017Braibant, 2019Hutsemekers, 2021Hutsemekers}, we show an example of two-dimensional distributions of the WCI and RBI indices. These indices, that characterize line profile distortions (Sect.~\ref{sec:bels}), were computed from the line profiles generated by the convolution of the KD, PW, and EW models with a magnification map. We see that a large diversity of WCI and RBI values, which encompass the observed ones, can be produced, even when considering a single BLR radius. The KD and EW models generate the strongest distortions, with comparable probabilities when combining all map orientations, and without fine-tuning. Finally, it is  worth noticing that the low magnification of the line core can be simultaneously fitted with the adopted KD and EW models, without assuming that the line core originates from a much larger, independent, BLR \citep{2024Fianc}.

To summarize, our results demonstrate that the assumption that ``size is everything'', which is useful to estimate the size and wavelength dependence of the smaller continuum source \citep{2005Mortonson}, cannot be applied to infer the BLR kinematics, without arbitrarily discarding some possible models.  We emphasize that our discussion refers to the microlensing of individual velocity bins, not to the estimation of the BLR size as a whole.  


\end{appendix}

\end{document}